\documentclass{IEEEtran}

\usepackage{hyperref}
\usepackage{caption}
\usepackage{indentfirst}

\usepackage{graphicx}
\usepackage{epstopdf}
\graphicspath{{fig/}}
\usepackage{subfigure}
\usepackage{float}
\usepackage{array}
\usepackage{url}
\usepackage{longtable}
\usepackage{rotating}
\usepackage{multirow}
\usepackage{amsmath}
\usepackage{amsfonts}
\usepackage{amssymb}
\usepackage{mathrsfs}
\usepackage{mdwlist}
\usepackage{booktabs}
\usepackage{threeparttable}
\usepackage{makecell}
\usepackage{diagbox}
\usepackage{pifont}

\usepackage{pdflscape}
\usepackage{footnote}

\usepackage[numbers,sort&compress]{natbib}
\usepackage[linesnumbered,boxed,ruled]{algorithm2e}
\usepackage[hang,flushmargin]{footmisc}
\usepackage{lipsum}

\usepackage[T1]{fontenc}
\usepackage[utf8]{inputenc}

\usepackage{color,xcolor}
\definecolor{orange}{RGB}{234,87,23}
\definecolor{mygreen}{RGB}{176,246,00}

\title{Bridging the Gap between Continuous and Informative Discrete Representations by Random Product Quantization}

\author{

    {Xueqing Li, Hao Ma, Zehan Li, Rujin Chen, Boyu Zhu, Ruihao Jing, Jian Kang, Jie Li,\\ Chi Zhang, Xiao-Lei Zhang, \IEEEmembership{Senior Member, IEEE}, and Xuelong Li, \IEEEmembership{Fellow, IEEE}}

    \thanks{Xiao-Lei Zhang is the corresponding author.}
    \thanks{Xueqing Li and Hao Ma contributed equally to this work.}
    \thanks{Xueqing Li, Hao Ma, Rujin Chen, Boyu Zhu, Ruihao Jing, and Xiao-Lei Zhang are with the School of Marine Science and Technology, Northwestern Polytechnical University, 127 Youyi West Road, Xi'an, Shaanxi 710072, China, with the Institute of Artificial Intelligence (TeleAI), China Telecom, and also with the Research and Development Institute of Northwestern Polytechnical University in Shenzhen, Shenzhen 518063, China (e-mail: \{lixueqing, haoma, chenrujin, zhuboyu, ruihaojing\}@mail.nwpu.edu.cn, xiaolei.zhang@nwpu.edu.cn).}
    \thanks{Zehan Li, Jian Kang, Jie Li, Chi Zhang, and Xuelong Li are with the Institute of Artificial Intelligence (TeleAI), China Telecom (e-mail: \{lizh85, kangj30, lij86, zhangc120\}@chinatelecom.cn, xuelong\_li@ieee.org ).}

}

\begin{document}
\maketitle

\begin{abstract}

    Self-supervised learning \textcolor{black}{(SSL)} has become a core technique in speech processing, but the high dimensionality of its representations makes discretization essential for improving efficiency. However, existing discretization methods still suffer from significant information loss, resulting in a notable performance gap compared to continuous representations. To overcome these limitations, we propose two quantization-based discretization methods: Product Quantization (PQ) and Random Product Quantization (RPQ). PQ partitions the original feature space into multiple subspaces and independently quantizes each sub-vector, producing a fused set of discrete units that retain diverse information from different subspaces\textcolor{black}{, thereby} mitigating the loss associated with single-cluster quantization. RPQ further enhances representation diversity by randomly sampling feature dimensions multiple times to construct sub-vectors, thereby better capturing the variability in the data distribution. Theoretical analysis shows that RPQ reduces the correlation $\rho\ (0 \le \rho \le 1)$ between sub-quantizers, and its quantization error is lower-bounded by $\rho \varepsilon_{\text{kms}}$, where $\varepsilon_{\text{kms}}$ is the error of a single K-means quantizer. Experimental results show that, on the combined dataset constructed from LibriSpeech and ML-SUPERB, PQ and RPQ outperform standard K-means discretization, achieving relative improvements of 21.8\% and 20.0\% in \textcolor{black}{word error rate (WER)} on LibriSpeech, and 24.1\% and 19.6\% in \textcolor{black}{character error rate (CER)} on the ML-SUPERB, respectively. Moreover, their performance is competitive with, and in some cases even surpasses, that of continuous SSL representations.
    
\end{abstract}

\section{Introduction} \label{sec:introduction}

\IEEEPARstart{I}{n} recent years, the rapid advancement of deep learning has greatly \textcolor{black}{improved} automatic speech recognition (ASR) \cite{abdel2012applying,graves2013speech}. Since the emergence of end-to-end ASR models, performance has been further improved with the aid of sufficient computational resources \cite{chan2016listen,vaswani2017attention,gulati2020conformer}. The performance of end-to-end ASR models depends on large amounts of transcribed data, but such data is limited and difficult to obtain. To alleviate this dependency, SSL has been widely adopted in the domain of ASR and has demonstrated remarkable success \cite{synnaeve2020end,kahn2020self}. During pre-training, SSL models learn meaningful speech representations from large amounts of unlabeled speech data \cite{schneider2019wav2vec,baevskivq,baevski2020wav2vec,hsu2021hubert,chen2022wavlm,baevski2022data2vec,baevski2023efficient}. In downstream training, the continuous speech representations generated by feeding raw waveforms into the pre-trained SSL model serve as inputs, resulting in better performance than using traditional acoustic features. Furthermore, techniques such as fine-tuning the pre-trained models or inserting adapters can further align SSL representations with specific downstream tasks.

Despite their effectiveness in improving downstream tasks, SSL representations impose significant storage demands and computational burdens due to their high-dimensional and continuous characteristics. Some studies have shown that the redundancy in high-dimensional acoustic features or SSL continuous representations can lead to inefficient sequence modeling \cite{chang2023exploration,chang2024exploring}. Consequently, recent work has explored using discrete units of speech representation in ASR tasks, where a discrete token from a restricted dictionary represents the speech signal within a short time frame. This method retains performance comparable to traditional acoustic features while significantly compressing the information and reducing data storage. For instance, \textcolor{black}{Chang \textit{et al.} \cite{chang2023exploration} cluster} high-dimensional continuous SSL representations into discrete units using K-means clustering, achieving approximately 3000-fold compression. Subsequently, \cite{chang2024exploring} handles temporal redundancy by applying de-duplication and subword modeling, further shortening the input discrete unit sequence length. In addition to significantly reducing model training and inference time, discrete speech representations can be viewed as a unique form of text representation in NLP. By employing discretized speech representations, the gap between speech processing tasks and the field of NLP can be effectively bridged. \textcolor{black}{This will enable large language models (LLMs) to receive both text tokens and speech tokens simultaneously during training, allowing the model to learn not only textual information but also the acoustic and semantic details within speech tokens. As a result, the model will be capable of effectively performing complex tasks that span across these two modalities \cite{nguyen2025spirit,huang2025step}.}


Current methods for speech representation discretization can be broadly categorized into two types. The first type involves training a neural codec \cite{van2017neural}, from which discrete speech tokens are obtained via its quantizer module. These tokens typically capture the physical properties of speech signals, such as speaker identity and prosody, and are thus referred to as acoustic tokens. The second type discretizes the output representations of SSL models to produce semantic tokens, which encode higher-level linguistic information such as word and syntactic content. This approach has been more widely adopted in ASR tasks and is the primary focus of this work. Specifically, some methods directly utilize built-in quantization modules in SSL models to generate discrete representations \cite{chiu2022self,liu2023dinosr,zhu2025muq}, while others apply K-means clustering to hidden representations from SSL models to extract discrete units \cite{chang2023exploration,chang2024exploring,cui2024exploring}. Since some SSL models do not include built-in quantization modules, applying external clustering methods to their representations offers broader applicability. However, most existing studies primarily focus on using K-means clustering to discretize SSL representations. While straightforward, this approach imposes a high degree of information compression, often leading to the loss of important semantic information. As a result, its performance on ASR tasks is typically inferior to that achieved with continuous SSL representations. More recently, residual vector quantization (RVQ) has been explored as an alternative to K-means \cite{shi2024mmm}. Although RVQ has shown potential, setting a large number of quantization layers may introduce instability due to the accumulation of high-order residuals, while using fewer layers often yields only marginal improvements.

To better preserve informative content during the discretization of SSL representations and narrow the performance gap between discrete and continuous representations, we propose two novel discretization methods based on Product Quantization (PQ) and Random Product Quantization (RPQ). We also conduct a thorough theoretical analysis of RPQ from the perspective of quantization error. PQ first divides continuous representations into multiple subspaces and independently quantizes each one. The discrete representation is then formed by a concatenation of the quantized indices from each subspace. Compared to \textcolor{black}{standard} K-means clustering, PQ can retain more useful information across multiple dimensions during compression, thereby reducing information loss more effectively. Building on PQ, RPQ further improves the subspace partitioning strategy by \textcolor{black}{performing multiple random selections of feature dimensions} to construct diverse low-dimensional sub-vectors. This enhances the diversity among subspaces and enables the discrete representation to capture the rich distribution of speech features more comprehensively. We conduct ASR experiments on a combined dataset of LibriSpeech and ML-SUPERB. Compared to standard K-means clustering, PQ improves performance by 21.8\% and 20.0\% on the two datasets. RPQ achieves even higher gains of 24.1\% and 19.6\%. Despite using much less training time than continuous SSL representations, the proposed methods achieve comparable or even better performance.

The contributions of this work can be summarized as follows:  
\begin{itemize}
    \item
    \textbf{We propose a speech representation discretization method based on PQ.} PQ divides continuous representations into multiple low-dimensional sub-vectors, which are then independently quantized and merged. Compared to the widely used K-means clustering, PQ effectively reduces information loss during the discretization process. 

    \item
    \textbf{We introduce RPQ for speech representation discretization.} RPQ constructs multiple low-dimensional sub-vectors by randomly selecting different subsets of dimensions from the continuous SSL feature space. This increases the diversity among sub-vectors and allows the resulting discrete representations to \textcolor{black}{more effectively} capture the rich distribution of speech features.

    \item
    \textbf{We conduct a thorough theoretical analysis of RPQ.} By decomposing the quantization error, we derive the error lower bound of RPQ under limiting conditions. This analysis also provides guidance for parameter selection in RPQ.
\end{itemize}

In the following, Section \ref{related} categorizes discrete representations into two types and reviews related work for each. Section \ref{sec:framework} describes the speech recognition framework based on the discrete representation. Section \ref{propose} describes our proposed representation discretization based on PQ and RPQ in detail and conducts a theoretical analysis of the proposed method. Section \ref{experiment} conducts experiments and analyzes the experimental results to validate the effectiveness of our proposed methods. Finally, section \ref{conclusion} summarizes the whole paper.

\section{Related Works} \label{related}

Speech discretization, also known as speech quantization, was originally designed for compressing speech signals by converting continuous speech representations into sequences of discrete integers, facilitating efficient storage and transmission in communication systems. In recent deep learning research, discrete speech tokens have been explored as intermediate representations for various speech processing tasks, demonstrating advantages in both storage and computational efficiency. Recent studies have extensively explored the development of speech discretization methods. Based on the type of information contained within speech tokens, they can be broadly categorized into acoustic tokens and semantic tokens, each derived from distinct underlying principles. \textcolor{black}{This section briefly reviews the discretization methods for these two types of speech tokens. In particular, we also survey recent work on discrete SSL tokens for spoken language modeling (SLM).}

\subsection{Acoustic Token}

Acoustic tokens primarily capture the physical characteristics of speech signals, including pitch, pronunciation, and rhythm. Acoustic tokens are derived from \textcolor{black}{audio codecs}, which is a signal processing method used for compression and reconstruction \cite{guo2025recent}. Its primary goal is to reduce the bitrate while preserving the quality of the original signal as much as possible. In traditional audio processing, \textcolor{black}{audio codecs} is widely used for compressing and storing audio content such as music and films. The goal is to minimize storage requirements and transmission bandwidth while maintaining auditory quality, typically categorized into lossy \cite{finlayson2008more} and lossless compression. In speech processing, speech codec focuses on low-bitrate speech encoding, as seen in EVS \cite{dietz2015overview} for high-definition voice communication and Opus \cite{valin2012definition} for VoIP calls.

Neural codecs leverage deep learning for audio encoding and decoding. Through end-to-end training, they optimize the compression and reconstruction process and have attracted significant attention in recent years \cite{defossez2022high,jiang2023disentangled,kumar2023high,guo2024lscodec,shi2024espnet,zhang2024high}. Neural codecs are extensively applied in tasks such as text-to-speech synthesis (TTS), speech-to-speech translation (S2ST), and music generation. A typical neural codec model follows an Encoder-Quantizer-Decoder structure: the encoder extracts a compact representation of the raw audio, the quantizer converts the continuous feature vectors from the encoder into discrete tokens to reduce the bitrate of data representation, and the decoder reconstructs the audio signal from the quantized tokens while minimizing information loss \cite{van2017neural}. Many online learnable vector quantization methods have been incorporated into the quantizer module of neural codec models, including Gumbel-Softmax quantization \cite{jang2017categorical,baevski2019vq}, Finite Scalar Quantization \cite{mentzer2024finite, parker2025scaling}, and RVQ \cite{defossez2022high,kumar2023high}.

\subsection{Semantic Token}

Semantic tokens typically refer to discrete tokens derived from SSL representations. These representations capture not only the linguistic meaning of speech content but also higher-level information such as syntax, vocabulary, and semantic structure. SSL representations have been shown to significantly outperform traditional acoustic features in various downstream tasks. Discrete SSL representations help reduce storage and computational costs, and some studies suggest that discrete tokens can enhance speaker privacy protection \cite{chang2023exploration}. Semantic tokens are primarily used as inputs for recognition-based downstream tasks, such as ASR \cite{chang2023exploration, chang2024exploring, cui2024exploring,cui2024exploring_2, shi2024mmm} and speech translation. More recently, research has also explored their applications in generative tasks \cite{du22b_interspeech}.

SSL enables models to learn meaningful speech representations directly from raw audio by designing pretext tasks \cite{mohamed2022self}. There are two main approaches to obtaining discrete SSL tokens: (1) \textit{Internal quantization}: In SSL models that incorporate an internal quantizer module \cite{chiu2022self,liu2023dinosr,zhu2025muq}, the output of this quantizer can be directly used as semantic tokens. For example, VQ-wav2vec \cite{baevskivq} employs a vector quantization module to discretize continuous speech features into tokens, which serve as targets for the pretraining task. Very recently, DQ-Data2vec \cite{shao2025dq} introduces two decoupled quantizers to extract phoneme-level and language-level representations.

(2) \textit{External quantization}: The most common method involves training an additional clustering model to quantize the continuous representations from one or multiple layers of a pretrained SSL model into discrete units \cite{chang2023exploration, chang2024exploring, cui2024exploring}. Some studies also train dedicated codecs for tokenizing semantic representations \cite{huang2024repcodec,wang2024maskgct}. Existing external quantization methods are relatively limited in diversity, with most approaches relying on K-means clustering to generate semantic tokens. However, the high dimensionality of continuous SSL representations leads to increased computational cost and significant loss of detailed speech information during quantization \cite{guo2025recent}. \textcolor{black}{Recent advances have explored various strategies to improve external quantization: preprocessing SSL features with independent component analysis (ICA) before K-means clustering \cite{nakamura2025discrete}, learning optimal weighted combinations of multi-layer SSL representations \cite{li2025multilingual}, and enabling end-to-end joint optimization of tokenization and downstream tasks through differentiable K-means \cite{onda2025differentiable}.} The proposed PQ and RPQ methods effectively address these limitations and offer new insights into the discretization of SSL representations.

\subsection{\textcolor{black}{Prior Work on Discrete SSL Tokens for SLM}}

\textcolor{black}{The core goal of SLM is to learn the generative structure of language from raw audio without relying on text annotations. One significant milestone in introducing discrete tokens to the SLM field is the GSLM framework \cite{lakhotia2021generative}, which proposed a complete unsupervised pipeline: discretizing continuous SSL features into discrete units and training a language model on these units. Subsequent research \cite{nguyen2022discrete} demonstrated that discretization is crucial for SLM, as it helps eliminate acoustic redundancies (e.g., speaker differences, background noise), enabling the language model to focus more on syntax and semantics. However, discretization presents two key challenges: limited token granularity and inefficient information retention. A comparative study \cite{wang2024comparative} demonstrated that when discrete units are used in speech LLMs, information loss becomes particularly noticeable in higher-order semantic tasks (such as dialogue understanding), while continuous features are more easily understood by LLMs. To address this issue, Shon \textit{et al.} \cite{shon2024discreteslu} explored efficient fusion of discrete units with LLMs, proposing full-parameter fine-tuning of LLMs and designing lightweight adapters to align discrete units with the embedding space of LLMs, improving the effectiveness of discrete units, especially in speech understanding tasks, with better noise robustness and data efficiency. These studies collectively show that the performance of SLMs is closely tied to the quality of discrete units. The discretization method proposed in this paper enhances the information richness of discrete units and can seamlessly integrate with existing SLM frameworks.}

\section{Framework of ASR based on discrete representation}\label{sec:framework}

\begin{figure}
    \centering
    \includegraphics[width=1.0\linewidth]{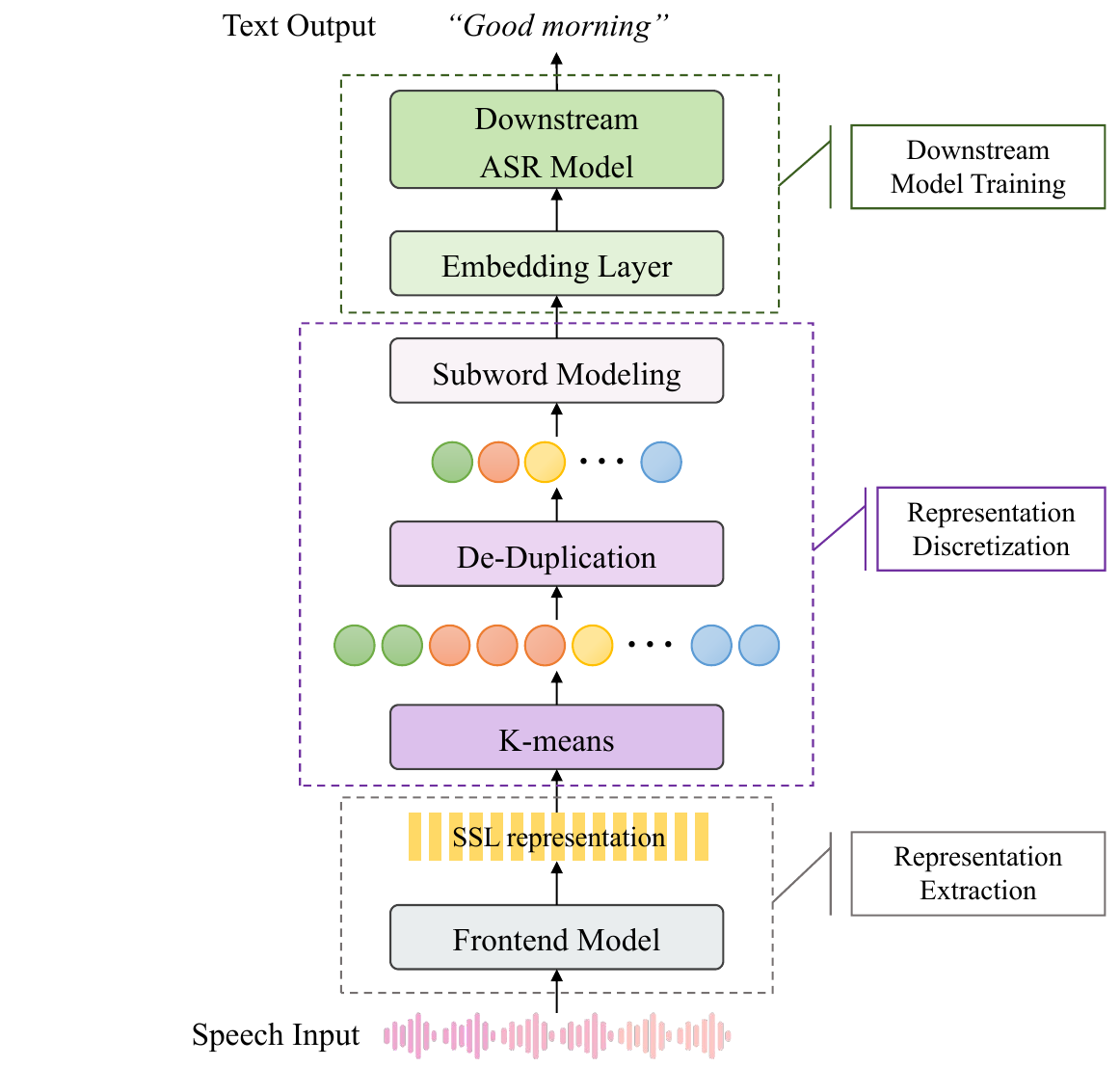}
    \caption{Framework of ASR model based on discrete representation.}
    \label{fig:fig-1}
\end{figure}

Fig. \ref{fig:fig-1} illustrates the baseline framework of the ASR model based on discretized SSL semantic tokens. As shown in the figure, the framework consists of three main components: self-supervised representation extraction, representation discretization, and downstream ASR model training. Given an input audio sequence represented as $ \boldsymbol{s}=\left[ s_1, s_2, \dots, s_L \right] $, where $ L $ denotes the number of frames in the audio and $ s_i $ represents the $ i $th frame: 

1) A pretrained SSL model is first used to extract the continuous semantic representations of $ \boldsymbol{s} $, denoted as $ \boldsymbol{x}=\left[ x_1, x_2, \dots, x_L \right] $, where each $ x_i $ is a $ D $-dimensional vector representing the hidden feature of the $ i $th audio frame.  

2) A trained K-means model with $ K $ clusters is then applied to discretize $ \boldsymbol{x} $ into a sequence of discrete units $ \boldsymbol{u}=\left[ u_1, u_2, \dots, u_L \right] $. The resulting discrete sequence maintains the same length as the hidden representations from the SSL model, where each $ u_t $ is obtained by minimizing the Euclidean distance between the feature $ x_t $ and the closest centroid $ c_u $:
\begin{equation}
    u_t={\arg\min}_{u\in 1,\cdots ,K}\lVert x_t-c_u \rVert ^2.
\end{equation}
After the aforementioned processing, the sequence of discrete tokens remains temporally aligned with the original speech features. However, these tokens still contain redundancies, such as repeated or co-occurring units. Notably, once speech signals are converted into discrete units, they can be viewed as a special type of language, similar to tokenized text in traditional NLP tasks. This enables the direct application of established NLP techniques for text processing and modeling. To reduce redundancy caused by repetition or co-occurring units and speed up training, following the settings in \cite{chang2024exploring}, we employ deduplication and subword modeling to shorten the input sequence. Deduplication compresses consecutive identical tokens into a single token. Subword modeling iteratively merges the two most frequent consecutive tokens and adds the merged token to the vocabulary \cite{guo2025recent}.

3) Finally, the processed discrete unit sequences serve as inputs to the downstream ASR model, where the corresponding transcription text acts as the learning target for the ASR task.

\section{Discretize based on Random Product Quantization} \label{propose}

\subsection{Product Quantization}

Considering that K-means clustering\textcolor{black}{, which maps $D$-dimensional continuous representations to discrete codebook indices, may incur excessive information loss, we propose using product quantization to improve representation discretization.} Vector quantization is typically used for lossy data compression, functioning by encoding values in a multidimensional vector space into a finite set of values in a lower-dimensional discrete subspace. \textcolor{black}{We optimize K-means clustering by using PQ, an efficient vector quantization technique introduced by Jégou \textit{et al.} \cite{jegou2010product}. This method aims to solve the problem of approximate nearest neighbor search in high-dimensional data, achieving memory optimization and speed improvement. PQ decomposes each original vector into several lower-dimensional subvectors and independently quantizes each subvector; the full codebook of PQ is then the Cartesian product of the sub-codebooks.} In this way, each vector can be represented by a combination of quantized codes from multiple low-dimensional spaces.

\begin{figure}
    \centering
    \includegraphics[width=0.9\linewidth]{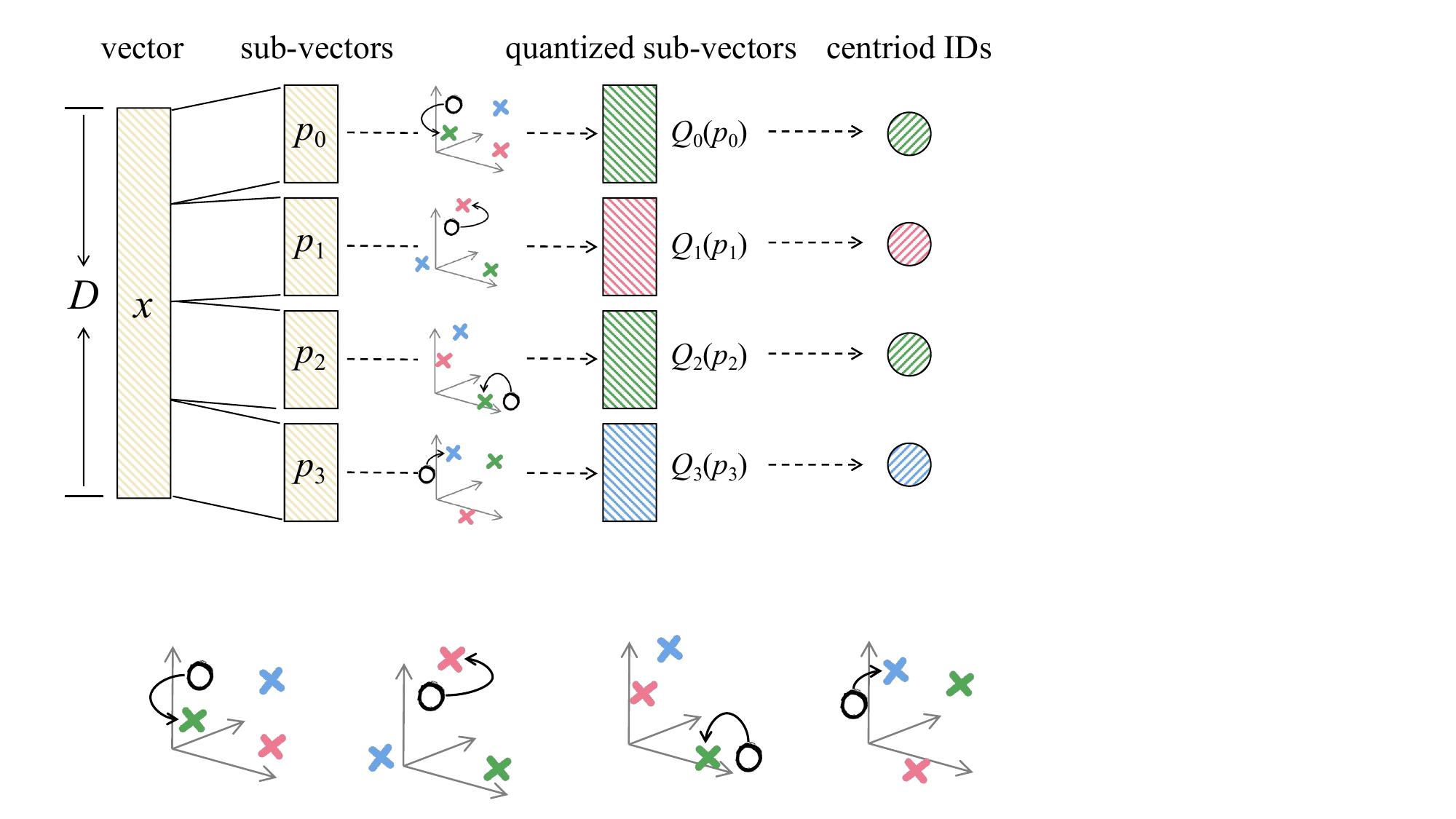}
    \caption{Instruction of PQ, taking $M=4$ as an example, where $Q_m\left( \cdot \right) $ represents the sub-quantizer corresponding to the sub-vector $p_m$.}
    \label{fig.pq}
\end{figure}

\begin{figure}
    \centering
    \includegraphics[width=1\linewidth]{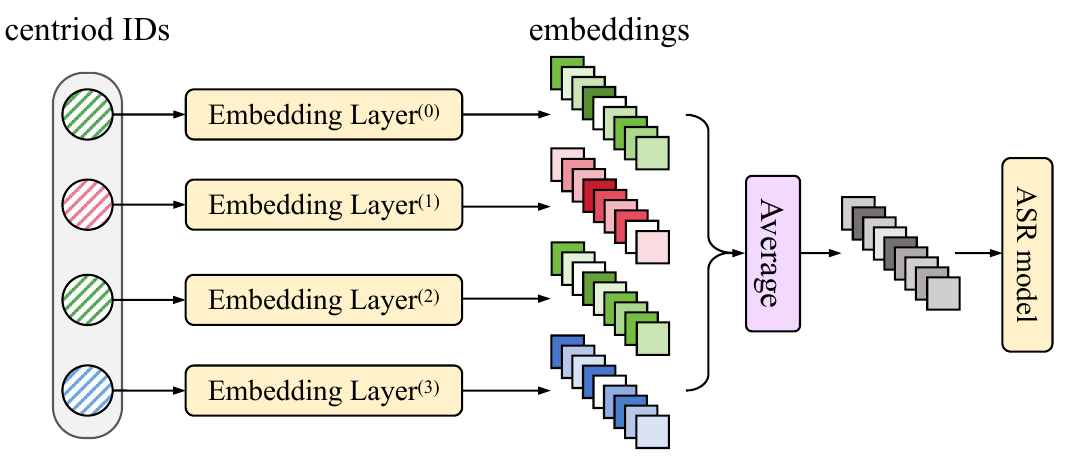}
    \caption{Illustration of discrete token merging.}
    \label{fig.merge}
\end{figure}

For the convenience of description, we use $x$ to represent $X_i$ in the following text, denoting the hidden embedding of a certain frame of input, where $x \in \mathbb{R}^D$. In typical vector quantization, the quantizer can be viewed as a function $Q$ that maps $x$ to a vector $ Q\left( x \right) \in \mathcal{C}=\left\{ c_1,c_2,\cdots,c_K \right\} $, where $\mathcal{C}$ represents the quantization codebook, $c_i$ is the centroid in the codebook $\mathcal{C}$, and $K$ is the size of the codebook, i.e., the number of centroids. In PQ, the input vector $x$ is equally split into $M$ sub-vectors $p_m \in \mathbb{R}^{d}$, where $0 \le m \le M-1$, $d = D/M$, and $D$ is divisible by $M$. Then $M$ different sub-quantizers $Q_m$ map these sub-vectors to their nearest centroids, as shown in Fig. \ref{fig.pq}. Each sub-quantizer correspondingly \textcolor{black}{$Q_m$ is associated with} its own codebook $\mathcal{C}_m$. The overall codebook of the product quantizer is defined as the Cartesian product of the codebooks of the $M$ sub-quantizers: $\mathcal{C}=\mathcal{C}_0\times \mathcal{C}_1\times \cdots \mathcal{C}_{M-1}$. Consequently, the centroids of the complete codebook are the concatenations of the centroids of the $M$ sub-quantizers. In the experiments presented in this paper, all sub-quantizers utilize the same number of centroids $k^*$. Therefore, the number of centroids of the PQ is $k=\left( k^* \right) ^M$. Clearly, explicitly storing all $k$ centroids is inefficient; in practice, PQ stores only the centroids of the sub-quantizers, which amounts to $M\times k^*$ centroids. Additionally, as illustrated in Fig. \ref{fig.pq}, we replace the quantized $d$-dimensional vectors with the indices of the centroids. Therefore, a $D$-dimensional vector $x$ is ultimately discretized into $M$ integer index tokens.

Since the continuous input representation at each time step is quantized into discrete tokens across multiple subspaces, these tokens should be merged into a format compatible with the downstream ASR model\textcolor{black}{—specifically, preserving temporal alignment and producing a single embedding per time step. To preserve alignment, we omit the de-duplication and subword modeling steps, since these operations typically alter token timing and sequence length. Concretely, as shown in Fig. \ref{fig.merge}, each of the $M$ indices is mapped via its own embedding table; these $M$ embeddings are then element-wise averaged to produce the final input embedding supplied to the ASR encoder.}

\subsection{Random Product Quantization}

\begin{figure}
    \centering
    \includegraphics[width=1.0\linewidth]{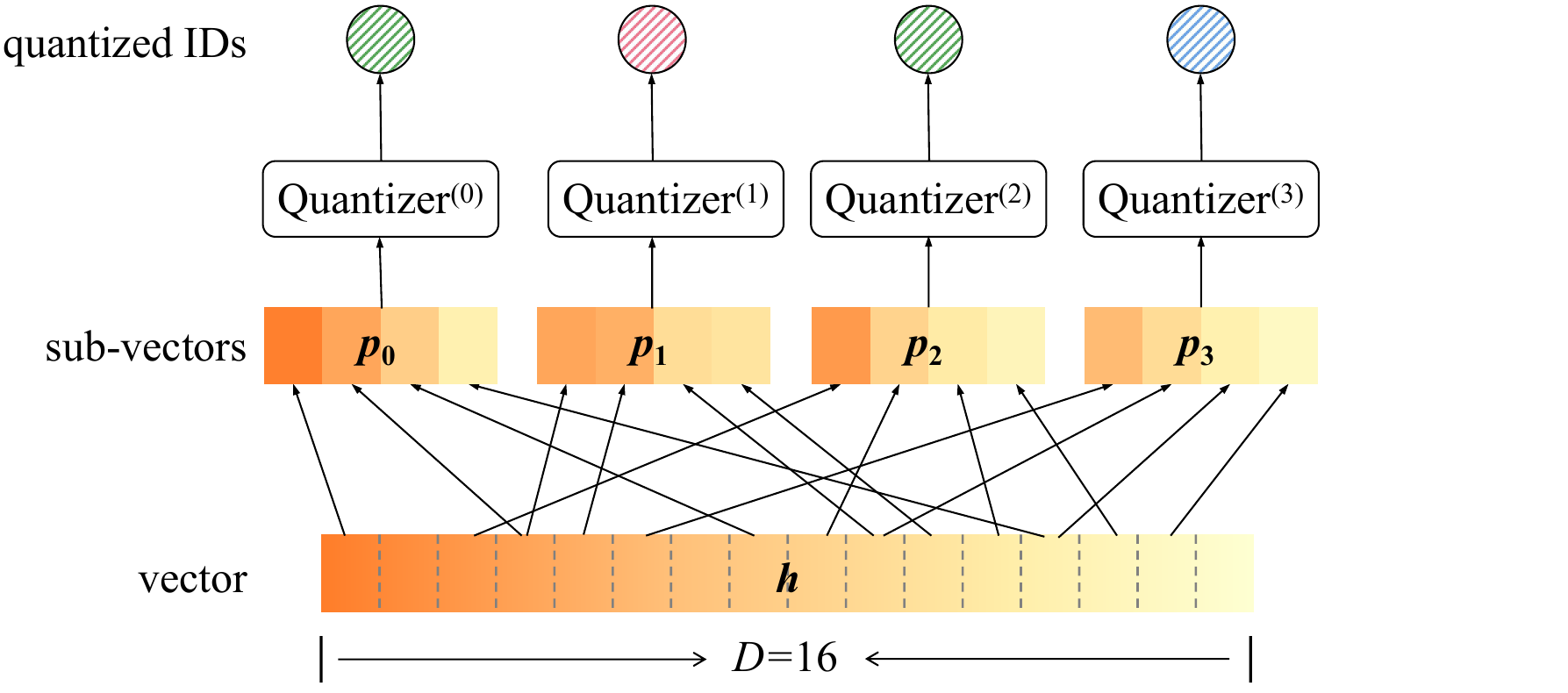}
    \caption{Instruction of RPQ, taking $D=16, M=4, \alpha=25\%$ as an example.}
    \label{fig.rpq}
\end{figure}

PQ partitions the original vector space into several subspaces by non-overlapping segmentation of the original vector. To enable the quantized discrete tokens to retain more comprehensive information from the continuous representation, we further propose RPQ. In RPQ, the segmentation rule is optimized to perform multiple random samplings within the feature space. The number of these random samplings can be considered the number of subspaces formed under the new rule. As shown in Fig. \ref{fig.rpq}, we randomly extract $M$ sub-vectors of dimension $d = \alpha \times D$ from the input vector $x$ according to a predefined proportion $\alpha$ \((0 < \alpha \leq 1)\), forming a set of sub-vectors $\{ p_0, p_1, \dots, p_{M-1} \}$. Subsequently, we train $M$ corresponding sub-quantizers to quantize these sub-vectors into discrete tokens, and we use the same merging method as in PQ to fuse multiple tokens before feeding them into the encoder of the downstream ASR model.

\subsection{Theoretical Analysis}

This subsection provides a theoretical analysis of discretization methods using PQ and RPQ from the perspective of ensemble learning. Additionally, it examines the memory efficiency of PQ. 

The high-dimensional continuous representations derived from self-supervised models contain rich speech attributes, including pronunciation features, speaking rate, and semantics. A single token extracted by K-means is \textcolor{black}{insufficient} to represent the complex characteristics of speech, and such compression leads to information \textcolor{black}{loss}, which limits downstream performance. In PQ, on the other hand, the representation is segmented into sub-vectors, enabling each sub-vector to focus on describing some specific attributes. By discretizing each sub-vector independently, the speech information of each frame can be preserved with greater detail. Compared to a single K-means clustering, a set of K-means clustering models generates multiple tokens for each frame, providing the ability to discretize the representation of a frame in more dimensions. The downstream model can leverage the combined information of multiple tokens to enhance the accuracy of speech recognition.

The discretization method in RPQ is inspired by \textcolor{black}{bootstrap aggregating (Bagging)}, a classical ensemble method that trains multiple models on different data subsets to reduce variance. The idea is to train multiple models on different subsets of the data and then combine their predictions to get a stronger model. This is done through Bootstrap sampling, which is a method of sampling with replacement to create different subsets of the original dataset. RPQ adopts the concept of Bootstrap sampling by randomly selecting a certain proportion of dimensions to form multiple subspaces. Each subspace can be regarded as a random sampling of the original feature space, ensuring that different K-means models are trained on distinct subspaces, which increases the diversity of the models. Furthermore, during training, they essentially cluster different feature dimensions, thereby capturing distinct data distributions.

It is important to note that RPQ is not the same as the resampling in Bagging. The core of Bagging lies in resampling the data, where multiple sub-datasets are created by randomly selecting samples from the dataset. In contrast, RPQ involves resampling the feature space, where different dimensions are randomly selected to construct subspaces. This approach does not involve data resampling, but instead increases model diversity by selecting different features.

\begin{figure}
    \centering
    \includegraphics[width=0.57\linewidth]{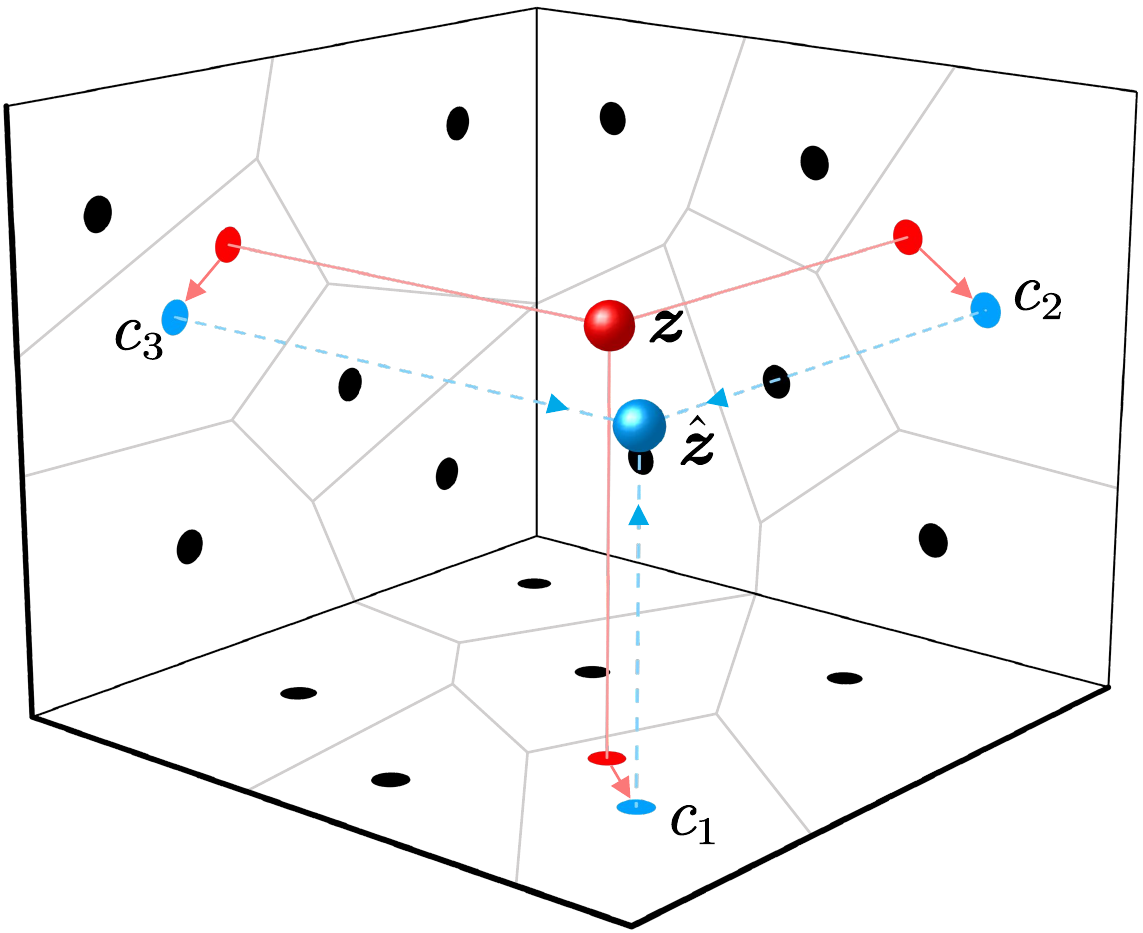}
    \caption{Illustration of the theoretical analysis of RPQ quantization error. For clarity, we take $M=3$ as an example, where the three planes represent three subspaces.}
    \label{fig.theory}
\end{figure}

To better analyze the effectiveness of RPQ, in the following we provide a theoretical analysis of the estimation error introduced during the quantization process. For a feature vector $\boldsymbol{x}$ at a specific time frame, $d$ dimensions are randomly selected $M$ times from the original feature space $\mathbb{R}^D$, forming $M$ subspaces $\left\{ \mathbb{R}_{m}^{d} \right\} _{m=1}^{M}$. Assume that the true local coordinate of $\boldsymbol{x}$ in the reconstructed subspace formed by these subspaces is $\boldsymbol{z}$, which is a fixed point in the vicinity of $\boldsymbol{x}$, as shown in Fig. \ref{fig.theory}. \textcolor{black}{Formally, we define $\boldsymbol{z}$ as the orthogonal projection of $\boldsymbol{x}$ onto the locally linear subspace reconstructed from $\left\{ \mathbb{R}_{m}^{d} \right\} _{m=1}^{M}$.} Theoretically, when the density of the nearest centroids $\left\{ \boldsymbol{c}_m \right\} _{m=1}^{M}$ around $\boldsymbol{x}$ in each subspace approaches infinity, $\boldsymbol{z}$ can be determined. In practice, however, when the number of nearest centroids around $\boldsymbol{x}$ is finite, we assume that $\boldsymbol{x}$ is projected to an estimated coordinate $\boldsymbol{\hat{z}}$. The effectiveness of the RPQ can be evaluated by estimating the error between $\boldsymbol{\hat{z}}$ and the true coordinate $\boldsymbol{z}$. Formally, the estimation error is defined as:
\begin{equation}
    \mathbb{E}\left( \lVert \boldsymbol{z}-\boldsymbol{\hat{z}} \rVert ^2 \right),
\end{equation}
where $\mathbb{E}\left( \cdot \right) $ represents the expectation operator. Analyzing this error is essential for evaluating RPQ's ability to preserve the information of the original continuous representations during the discretization process.

Assume that the sub-vectors used to train multiple K-means clustering models in RPQ are identically distributed but not necessarily independent, with a positive correlation coefficient  $\rho \ \left( 0\le \rho \le 1 \right) $ between the sub-vectors of two subspaces. Consequently, the correlation coefficient between the centroids of different subspaces (denoted as $\left\{ \boldsymbol{c}_{m_1}^{i} \right\} _{i=1}^{k^*} $ and $\left\{ \boldsymbol{c}_{m_2}^{j} \right\} _{j=1}^{k^*}$, where $\forall m_1,m_2=1,\cdots, M $ and $m_1\ne m_2$) is also $\rho$. \textcolor{black}{Let $\mathbb{S}_m$ denote the local neighborhood (cluster region) associated with a centroid in subspace $m$. If the number of clusters $k^*$ is large (approaching the number of input samples), these neighborhoods shrink and local linearity becomes a valid approximation within each $\mathbb{S}_m$.} 

\textcolor{black}{Since the dimensions of SSL features are not completely independent, we introduce a covariance matrix $\Phi$ to model the correlation between the dimensions. This matrix captures the relationships between the feature dimensions. Let the original SSL feature vector be $\boldsymbol{x} = [x_1, x_2, \dots, x_D]^T$, then $\Phi$ is defined as:}
\begin{equation}
    \textcolor{black}{
        \Phi = \begin{bmatrix}
        \varphi_{11} & \varphi_{12} & \cdots & \varphi_{1D} \\
        \varphi_{21} & \varphi_{22} & \cdots & \varphi_{2D} \\
        \vdots & \vdots & \ddots & \vdots \\
        \varphi_{D1} & \varphi_{D2} & \cdots & \varphi_{DD}
        \end{bmatrix},
    }
\end{equation}
\textcolor{black}{where $\varphi_{ij}$ represents the covariance between $x_i$ and $x_j$. To simplify the derivation, we assume that the diagonal elements $\varphi_{ii}$ of the covariance matrix represent the variance of each feature, and the off-diagonal elements represent the covariance between different features. This covariance matrix $\Phi$ is used to model the correlation between the feature dimensions.}

\begin{figure}
    \centering
    \includegraphics[width=0.65\linewidth]{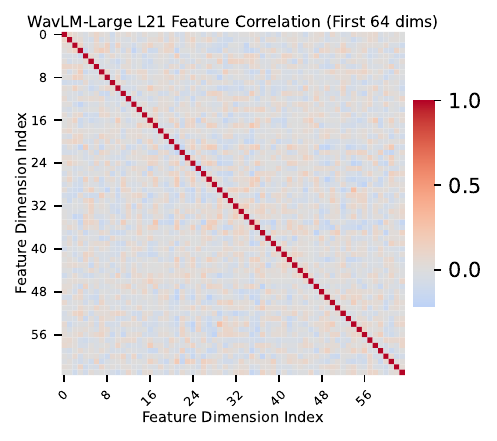}
    \caption{Heatmap of correlation between feature dimensions of WavLM Large Layer 21.}
    \label{fig.heatmap}
\end{figure}

Under these conditions, the estimation error of RPQ can be decomposed as follows:
\begin{equation}
    \mathbb{E}\left( \lVert \boldsymbol{z}-\boldsymbol{\hat{z}} \rVert ^2 \right) =\sum_{i=1}^d{\mathbb{E}\big( \left( z_i-\hat{z}_i \right) ^2 \big)}  \textcolor{black}{+ \sum_{i \neq j} \mathbb{E}(\epsilon_{ij})},
\end{equation}
where $\boldsymbol{z} = [z_1, \dots, z_d]^T$ and $\boldsymbol{\hat{z}} = [\hat{z}_1, \dots, \hat{z}_d]^T$. \textcolor{black}{Additionally, $\boldsymbol{\epsilon} = \Phi - \text{diag}(\Phi)$, where $\text{diag}(\Phi)$ extracts the diagonal part of the covariance matrix (i.e., the variances of each feature), and $\boldsymbol{\epsilon}$ represents the correlation between the dimensions. To assess the actual correlation between the feature dimensions, a heatmap was generated to visualize the pairwise correlations. The heatmap\footnote{\color{black} To enhance visibility, we only show correlations between the first 64 dimensions. The full heatmap can be retrieved on \href{https://aisaka0v0.github.io/rpq_supplementary_page/}{https://aisaka0v0.github.io/\\rpq\_supplementary\_page/}.}, shown as Fig. \ref{fig.heatmap}, indicates that the correlations between most feature dimensions are relatively weak. Based on this observation, and for the sake of simplifying the subsequent theoretical derivations, the error term $\mathbb{E}(\epsilon_{ij})$ will be excluded from further consideration.}

we next analyze $\mathbb{E}\big( \left( z_i - \hat{z}_i \right)^2 \big)$ in a single dimension, denoted as $\mathbb{E}\big( \left( z-\hat{z} \right) ^2 \big) $ for brevity. Under the assumption of local linearity, we use the deviation-variance decomposition formula of expected generalization error to decompose the estimation error into the following form:
\begin{equation}
    \begin{aligned}
	\mathbb{E}\big( \left( z-\hat{z} \right) ^2 \big) &=\mathbb{E}\big( \left( z-\mathbb{E}\left( \hat{z} \right) +\mathbb{E}\left( \hat{z} \right) -\hat{z} \right) ^2 \big)\\
	&=\left( z-\mathbb{E}\left( \hat{z} \right) \right) ^2+\mathbb{E}\big( \left( \hat{z}-\mathbb{E}\left( \hat{z} \right) \right) ^2 \big).\\
    \end{aligned}
\end{equation}
\textcolor{black}{Here, $z$ represents the scalar component of the multi-dimensional vector $\boldsymbol{z}$, which is a fixed value. $\hat{z}$ denotes the estimated value of $z$, and it's a random variable. $\mathbb{E}(\cdot)$ denotes the deterministic expectation of a random variable.}

The difference between the expected output and the true coordinates represents the bias, given by:
\begin{equation}
    \text{Bias}^2\left( \hat{z} \right) =\left( z-\mathbb{E}\left( \hat{z} \right) \right)^2.
    \label{eq.bias}
\end{equation}
The variance generated during the training phase is expressed as:
\begin{equation}
    \text{Var}\left( \hat{z} \right) =\mathbb{E}\big( \left( \hat{z}-\mathbb{E}\left( \hat{z} \right) \right) ^2 \big).
    \label{eq.var}
\end{equation}
Thus, $\mathbb{E}\big( \left( z-\hat{z} \right) ^2 \big) =\text{Bias}^2\left( \hat{z} \right) +\text{Var}\left( \hat{z} \right)$.

Under the assumption of local linearity, $\boldsymbol{c}_m$ can be further assumed to follow a multivariate normal distribution centered around $\boldsymbol{z}$. For each single dimension, $c_m$ follows a univariate normal distribution. Assuming that the variance of this normal distribution is $\sigma ^2$, the following expressions can be derived:
\begin{equation}
    \begin{aligned}
	\mathbb{E}\left( c_m \right) &=z,\\
	\mathbb{E}\left( c_{m}^{2} \right) &=\sigma ^2+z^2,\\
	\mathbb{E}\left( c_{m_1}c_{m_2} \right) &=\rho \sigma ^2+z^2.\\
    \end{aligned}
    \label{eq.c_m}
\end{equation}

When only a single K-means is used for discretizing the representations, i.e., $M=1$, there is only one centroid $c_m$. In this case $\hat{z}=c_m$. Combining Eqs. \eqref{eq.bias} to \eqref{eq.c_m}, we have:
\begin{equation}
    \begin{aligned}
        \text{Bias}^2\left( \hat{z} \right) &=\left( z-\mathbb{E}\left( \hat{z} \right) \right) ^2=\left( z-\mathbb{E}\left( c_m \right) \right) =0,\\
        \text{Var}\left( \hat{z} \right) &=\mathbb{E}\big( \left( \hat{z}-\mathbb{E}\left( \hat{z} \right) \right) ^2 \big) \\
        &=\mathbb{E}\big( \left( c_m-\mathbb{E}\left( c_m \right) \right) ^2 \big) =\sigma ^2.\\
    \end{aligned}
\end{equation}
Thus, the estimation error of a single K-means is:
\begin{equation}
    \varepsilon _{\text{kms}}=\text{Bias}^2\left( \hat{z} \right) +\text{Var}\left( \hat{z} \right) =\sigma^2.
    \label{eq.kms}
\end{equation}
When RPQ is used, there is a set of centroids $\left\{ c_m \right\} _{m=1}^{M}$, in which case the expected output is:
\begin{equation}
    \hat{z}_{\varSigma}=\frac{1}{M}\sum\limits_{m=1}^M{c_m}.
    \label{eq.z_RPQ}
\end{equation}
Combining Eqs. \eqref{eq.bias} to \eqref{eq.c_m} and Eq. \eqref{eq.z_RPQ}, the bias and variance can be calculated as follows:
\begin{equation}
    \begin{aligned}
        \text{Bias}^2\left( \hat{z}_{\varSigma} \right) &=0, \\
        \text{Var}\left( \hat{z}_{\varSigma} \right) &= \text{Var}\left( \frac{1}{M}\sum\limits_{m=1}^M{c_m} \right) \\
        &=\left[ \frac{1}{M}+\left( 1-\frac{1}{M} \right) \rho \right] \sigma ^2.\\
        \end{aligned}
\end{equation}
Therefore, the estimation error of RPQ is:
\begin{equation}
    \begin{aligned}
    	\varepsilon_{\text{RPQ}}&=\text{Bias}^2\left( \hat{z}_{\Sigma} \right) +\text{Var}\left( \hat{z}_{\Sigma} \right)\\
    	&=\left[ \frac{1}{M}+\left( 1-\frac{1}{M} \right) \rho \right] \sigma^2.\\
    \end{aligned}
    \label{eq.RPQ}
\end{equation}
Based on Eqs. \eqref{eq.kms} and \eqref{eq.RPQ}, we can get the relationship between $\varepsilon _{\text{RPQ}}$ and $\varepsilon _{\text{kms}}$ as:
\begin{equation}
    \varepsilon _{\text{RPQ}}=\left[ \frac{1}{M}+\left( 1-\frac{1}{M} \right) \rho \right] \varepsilon _{\text{kms}}.
    \label{eq.RPQ_kms}
\end{equation}

Based on Eq. \eqref{eq.RPQ_kms}, we can get the following corollaries:
\begin{itemize}
    \item \textit{The estimation error $\varepsilon_{\text{RPQ}}$ of the RPQ discretization method is constantly smaller than the estimation error $\varepsilon_{\text{kms}}$ of a single K-means with $\varepsilon_{\text{kms}}$ as the upper limit. This proves the effectiveness of RPQ.}
    \item \textit{When the number of subspaces M tends to positive infinity, $\varepsilon_{\text{RPQ}}$ will reach the lower limit $\rho \varepsilon_{\text{kms}}$, and when $\rho$ decreases from 1 to 0, $\varepsilon _{\text{RPQ}}$ will decrease from $\varepsilon_{\text{kms}}$ down to $\varepsilon_{\text{kms}}/{M}$. \footnote{\textcolor{black}{Note that as $M$ increases, the newly formed subspaces tend to overlap more with existing ones, thereby providing less complementary information. Consequently, when $M$ becomes large, the performance gain of RPQ saturates, as shown in \autoref{tab:RVQ}.}}}

\end{itemize}

According to Eq. \eqref{eq.RPQ_kms}, it can be seen that $\varepsilon_{\text{RPQ}}$ is closely related to the correlation coefficient $\rho$. In the following, we focus on analyzing $\rho$ in detail. According to its definition, $\rho$ primarily depends on the overlap probability between the two subsets of $d$-dimensional features randomly selected from the total $D$-dimensional feature space. Suppose that the dimensions selected twice constitute dimension subsets $S_1$ and $S_2$, and $\left| S_1 \right|=\left| S_2 \right|=d$. The overlap between $S_1$ and $S_2$ is determined by the size of their intersection. For any given dimension, the probability of being selected is $d/D = \alpha$. Since the two selections are independent, the likelihood of the dimension being selected in both is $\left( d/D \right) ^2=\alpha ^2$. For $D$ total dimensions, the expectation of the number of overlapping dimensions is:
\begin{equation}
    \mathbb{E}\left( \left| S_1\cap S_2 \right| \right) =D\cdot \left( \frac{d}{D} \right) ^2=\frac{d^2}{D}.
\end{equation}
The similarity between $S_1$ and $S_2$ can be measured using the Jaccard similarity coefficient:
\begin{equation}
    \text{Jaccard}\left( S_1,S_2 \right) =\frac{\left| S_1\cap S_2 \right|}{\left| S_1\cup S_2 \right|}.
\end{equation}
Since $\left| S_1\cup S_2 \right| = 2d-\left| S_1\cap S_2 \right|$, the expected Jaccard similarity coefficient is:
\begin{equation}
    \begin{aligned}
        \mathbb{E}\left( \text{Jaccard}\left( S_1,S_2 \right) \right) &=\frac{\mathbb{E}\left( \left| S_1\cap S_2 \right| \right)}{\mathbb{E}\left( \left| S_1\cup S_2 \right| \right)} \\
        & = \frac{\frac{d^2}{D}}{2d-\frac{d^2}{D}}=\frac{\alpha}{2-\alpha}.\\
    \end{aligned}
\end{equation}
Since $\rho$ measures the correlation between two subspaces, the Jaccard similarity coefficient can indirectly reflect this correlation. Assuming that the correlation of sub-spaces is proportional to the similarity of their dimension subsets, the correlation coefficient $\rho$ can be approximated as:
\begin{equation}
    \rho =\frac{\alpha}{2-\alpha}.
    \label{eq.rho}
\end{equation}
Eq. \eqref{eq.rho} indicates that $\rho$ is linearly and positively correlated with $\alpha$. As $\alpha$ increases from 0 to 1, $\rho$ also increases from 0 to 1. Therefore, we can directly control the correlation coefficient $\rho$ by adjusting $\alpha$.

As indicated by Eqs. \eqref{eq.RPQ_kms} and \eqref{eq.rho}, increasing $M$ and decreasing $\alpha$ can help RPQ reduce estimation errors. However, the increase in $M$ in turn leads to computational complexity, so we need to choose an appropriate $M$ to achieve a balance between performance and computational complexity. Additionally, the decrease of $\alpha$ causes the increase of $\varepsilon_{\text{kms}}$, indicating that reducing $\varepsilon_{\text{kms}}$ and decreasing $\alpha$ are also conflicting factors. Thus, it is essential to set a suitable dimensional selection ratio $\alpha$.

\section{Experiments} \label{experiment}

\begin{table*}[htbp]
    \centering
    \caption{Performance w.r.t. different cluster center numbers $k$.}
      \begin{tabular}{cp{1.3cm}|cccccccccc}
      \toprule
      \multicolumn{2}{c}{\multirow{2}[4]{*}{\textbf{Method}}} & \multicolumn{2}{c}{\textbf{test-clean}} & \multicolumn{2}{c}{\textbf{test-other}} & \multicolumn{2}{c}{\textbf{dev-clean}} & \multicolumn{2}{c}{\textbf{dev-other}} & \multicolumn{2}{c}{\textbf{test-1h}} \\
      \cmidrule(lr){3-4} \cmidrule(lr){5-6} \cmidrule(lr){7-8} \cmidrule(lr){9-10} \cmidrule(lr){11-12}    \multicolumn{2}{c}{} & CER   & WER   & CER   & WER   & CER   & WER   & CER   & WER   & CER   & WER \\
      \midrule
      \multirow{4}[2]{*}{\textbf{K-means}} & $k$=500 & 2.1   & 5.9   & 4.3   & 10.7  & 2.1   & 6.0   & 4.3   & 10.3  & 24.9  & 69.6 \\
            & $k$=1000 & 1.7   & 5.0   & 3.7   & 9.2   & 1.8   & 5.1   & 3.4   & 8.6   & 24.1  & 68.4  \\
            & $k$=1500 & 1.7   &  4.8   & 3.4   & \textbf{8.5} & 1.8   & 4.9   & \textbf{3.3} & \textbf{8.2} & \textbf{23.7} & \textbf{67.9} \\
            & $k$=2000 & \textbf{1.5} & \textbf{4.6} & \textbf{3.3} & 8.6   & \textbf{1.6} & \textbf{4.7} & \textbf{3.3} & \textbf{8.2} & 24.0  & 68.3  \\
      \midrule
      \multirow{4}[2]{*}{\textcolor{black}{\textbf{PQ}}} & $k$=500 & \textcolor{black}{\bf 1.0}    & \textcolor{black}{\bf 3.4}      & \textcolor{black}{\bf 2.4}      & \textcolor{black}{6.9}      & \textcolor{black}{\bf 1.1}      & \textcolor{black}{\bf 3.4}      & \textcolor{black}{2.5}      & \textcolor{black}{6.7}      & \textcolor{black}{\bf 19.2}      & \textcolor{black}{59.2}     \\
            &  $k$=1000 & \textcolor{black}{1.1}    & \textcolor{black}{3.5}      & \textcolor{black}{\bf 2.4}      & \textcolor{black}{\bf 6.8}      & \textcolor{black}{\bf 1.1}      & \textcolor{black}{\bf 3.4}      & \textcolor{black}{\bf 2.4}      & \textcolor{black}{\bf 6.5}      & \textcolor{black}{19.4}      & \textcolor{black}{59.3}     \\
            & $k$=1500 & \textcolor{black}{1.1}    & \textcolor{black}{3.6}      & \textcolor{black}{2.5}      & \textcolor{black}{7.0}      & \textcolor{black}{\bf 1.1}      & \textcolor{black}{3.6}      & \textcolor{black}{2.5}      & \textcolor{black}{6.6}      & \textcolor{black}{19.4}      & \textcolor{black}{59.6}     \\
            & $k$=2000 & 1.1  & 3.7  & 2.5  & 7.0  & {\bf 1.1}  & 3.7  & 2.5  & 6.8  & 19.6  & {\bf 59.1} \\
      \midrule
      \multirow{4}[2]{*}{\textcolor{black}{\textbf{RPQ}}} & $k$=500 & \textcolor{black}{1.1}      & \textcolor{black}{3.5}      & \textcolor{black}{2.5}      & \textcolor{black}{6.8}      & \textcolor{black}{1.2}      & \textcolor{black}{3.5}      & \textcolor{black}{2.6}      & \textcolor{black}{\bf 6.4}      & \textcolor{black}{\bf 18.9}      & \textcolor{black}{\bf 58.4}     \\
            & $k$=1000 & \textcolor{black}{\bf 1.0}    & \textcolor{black}{\bf 3.3}      & \textcolor{black}{2.4}      & \textcolor{black}{\bf 6.7}      & \textcolor{black}{1.1}      & \textcolor{black}{3.5}      & \textcolor{black}{\bf 2.4}      & \textcolor{black}{\bf 6.4}      & \textcolor{black}{19.1}      & \textcolor{black}{58.6}     \\
            & $k$=1500 & \textcolor{black}{\bf 1.0}    & \textcolor{black}{\bf 3.3}      & \textcolor{black}{2.5}      & \textcolor{black}{6.8}      & \textcolor{black}{\bf 1.0}      & \textcolor{black}{\bf 3.3}      & \textcolor{black}{\bf 2.4}      & \textcolor{black}{\bf 6.4}      & \textcolor{black}{19.0}      & \textcolor{black}{58.5}     \\
            & $k$=2000 & {\bf 1.0} & 3.4 & {\bf 2.3} & 6.6 & {\bf 1.0} & 3.4 & {\bf 2.4} & {\bf 6.4} & 19.3  & 59.3 \\
      \bottomrule
      \end{tabular}%
    \label{tab:k}%
\end{table*}%

\subsection{Experimental Setup}

\textbf{Dataset:} The datasets used in this paper include LibriSpeech and ML-SUPERB. LibriSpeech\cite{panayotov2015librispeech} is a large English speech dataset commonly used in speech recognition research, containing 16 kHz sampled read speech. ML-SUPERB\cite{shi23g_interspeech} is a multilingual dataset covering approximately 140 languages, ranging from major languages to rare dialects. The training set consists of 100 hours of training data (train-clean-100) from LibriSpeech and about 220 hours of training data (train-1h) from ML-SUPERB. And the test set includes dev-clean, dev-other, test-clean, and test-other from LibriSpeech, as well as test-1h from ML-SUPERB. Additionally, to evaluate the proposed method on other datasets, experiments are conducted on LibriSpeech, ML-SUPERB, and a Chinese ASR dataset WenetSpeech\cite{zhang2022wenetspeech}, respectively.

\textbf{Feature Extraction:} We employ two models to extract speech representations. The first is the WavLM model\footnote{\url{https://github.com/microsoft/unilm/tree/master/wavlm}}, which is trained on English audio data \cite{chen2022wavlm}, and the second is the TeleSpeech-ASR model\footnote{\url{https://github.com/Tele-AI/TeleSpeech-ASR}}, which is based on the Data2vec2 architecture \cite{baevski2023efficient} and trained using MFCC features from Chinese dialects. For the LibriSpeech dataset, we extract representations from the 21st layer of WavLM. For the WenetSpeech dataset, we use the final layer of TeleSpeech-ASR to obtain representations. For the ML-SUPERB dataset, we extract representations from the corresponding layers of both models.

\textbf{Parameter Configuration of Discrete Representation:} The K-means models used in PQ and RPQ quantizers are trained on approximately 100 hours of data from the training set. Specifically, the proportion of data used for quantizer training is set to 30\% for the mixed dataset, 100\% for LibriSpeech, 50\% for ML-SUPERB, and 100\% for WenetSpeech. Unless otherwise specified, the number of cluster centers for both PQ and RPQ quantizers is fixed at 2000. To enhance randomness among multiple sub-quantizers, the initialization method for cluster centers in RPQ is set to \textit{random}. In the PQ and RPQ discretization experiments, redundant token removal and subword modeling are omitted for speech discretized sequences. For text output sequences, WenetSpeech, as a Chinese dataset, adopts \textit{char} as the subword modeling unit, while all other datasets use byte pair encoding (BPE) with a vocabulary size of 6000.

\textbf{Downstream Model:} Experiments are conducted using the open-source end-to-end speech processing toolkit ESPnet \footnote{\url{https://github.com/espnet/espnet}}. The downstream ASR model is based on E-Branchformer \cite{kim2023branchformer}. The learning rate is set to $5\times10^{-4}$ and adjusted dynamically using a warmup scheduler with the steps of 5k by default. The CTC weight is set to 0.3, and no external language models are used during training or decoding. For all experiments, the beam search size is fixed at 10.

\subsection{Implementation of Baseline and Analysis of Cluster Centers \texorpdfstring{$k$}{k}}

We first implement the baseline system using the K-means clustering-based discretization method. \autoref{tab:k} presents the speech recognition performance using different numbers of K-means cluster centers $k$. \textcolor{black}{Note that, unless otherwise specified, all K-means baselines in this paper incorporate deduplication and subword modeling.} As shown, both CER and WER gradually decrease as $k$ increases, reaching optimal performance at $k = 2000$. Based on this result, we set $k = 2000$ as the number of cluster centers in subsequent experiments.

\begin{table*}[htbp]
    \centering
    \tabcolsep=1.5pt
    \caption{Experimental results of self-supervised speech recognition based on discretization of PQ and RPQ representations.}
      \begin{tabular}{cp{1.3cm}|cccccccccc}
      \toprule
      \multicolumn{2}{c}{\multirow{2}[4]{*}{\textbf{Method}}} & \multicolumn{2}{c}{\textbf{test-clean}} & \multicolumn{2}{c}{\textbf{test-other}} & \multicolumn{2}{c}{\textbf{dev-clean}} & \multicolumn{2}{c}{\textbf{dev-other}} & \multicolumn{2}{c}{\textbf{test-1h}} \\
      \cmidrule(lr){3-4} \cmidrule(lr){5-6} \cmidrule(lr){7-8} \cmidrule(lr){9-10} \cmidrule(lr){11-12}   \multicolumn{2}{c}{} & CER  & WER  & CER  & WER  & CER  & WER  & CER  & WER  & CER  & WER  \\
      \midrule
      K-means &   & 1.5  & 4.6  & 3.3  & 8.6  & 1.6   & 4.7  & 3.3  & 8.2  & 24.0  & 68.3  \\
      \midrule
      Continuous SSL &   & 1.1  & 3.5  & 2.7  & 7.0  & 1.4  & 3.7  & 2.5  & \textbf{6.4} & 21.7  & 63.9  \\
      \midrule
      \multirow{5}[2]{*}{\textbf{PQ}} & $M$=2 & 1.2  & 3.9  & 2.7  & 7.5  & 1.3  & 3.9  & 2.7  & 7.2  & 21.8  & 64.6  \\
            & $M$=4 & 1.1  & 3.7  & 2.6  & 7.2  & 1.2  & 3.7  & 2.6  & 6.9  & 20.0  & 60.3 \\
            & $M$=8 & 1.1  & 3.5  & 2.4  & 6.8  & 1.1  & \textbf{3.4} & 2.5  & 6.6  & 19.5  & 59.7 \\
            & $M$=16 & 1.1  & 3.5  & 2.4  & 6.9  & 1.1  & 3.5  & \textbf{2.4} & 6.5  & \textbf{19.2} & \textbf{58.9} \\
            & $M$=32 & 1.1  & 3.7  & 2.5  & 7.0  & 1.1  & 3.7  & 2.5  & 6.8  & 19.6  & 59.1 \\
      \midrule
      \multirow{5}[2]{*}{\textbf{RPQ}} & $M$=2 & 1.3  & 4.0  & 2.9  & 7.9  & 1.3  & 3.9  & 2.9  & 7.5  & 21.4  & 63.6 \\
            & $M$=4 & 1.1  & 3.7  & 2.6  & 7.3  & 1.2  & 3.7  & 2.7  & 7.1  & 20.4  & 61.3 \\
            & $M$=8 & 1.1  & 3.6  & 2.5  & 7.0  & 1.2  & 3.6  & 2.6  & 6.7  & 19.9  & 60.4 \\
            & $M$=16 & \textbf{1.0}  & \textbf{3.4}  & 2.4  & 6.8  & 1.1  & 3.5  & 2.5  & 6.7  & 19.5  & 59.6 \\
            & $M$=32 & \textbf{1.0} & \textbf{3.4} & \textbf{2.3} & \textbf{6.6} & \textbf{1.0} & \textbf{3.4} & \textbf{2.4} & \textbf{6.4} & 19.3  & 59.3 \\
            \midrule
            \multicolumn{2}{c}{\color{black}{\bf RPQ}, $M$=32} {\color{black}\it w/ varing ramdom seeds} & {\color{black}1.05$\pm$0.06} & {\color{black}3.43$\pm$0.05} & {\color{black}2.38$\pm$0.05} & {\color{black}6.70$\pm$0.08} & {\color{black}1.05$\pm$0.06} & {\color{black}3.43$\pm$0.05} & {\color{black}2.40$\pm$0.00} & {\color{black}6.45$\pm$0.06} & {\color{black}19.40$\pm$0.08}  & {\color{black}59.40$\pm$0.08}\\
      \bottomrule
      \end{tabular}%
    \label{tab:main}%
\end{table*}

\subsection{Experimental Results of PQ and RPQ}

\autoref{tab:main} compares the representation discretization methods based on K-means, PQ, and RPQ, alongside results obtained using continuous SSL representations. 
{
\color{black}
We further analyze the parameter sensitivity of PQ and RPQ with respect to the number of cluster centers $k$. Unlike the k-means baseline, PQ and RPQ employ a multi-subspace clustering strategy, under which the optimal number of cluster centers $k$ per subspace tends to be smaller than that of the global clustering baseline (k-means) for achieving the best performance.
Nevertheless, for fair comparison, we fix $k=2000$ as the standard experimental setting in the subsequent experiments.
}
Compared to the baseline method that utilizes K-means for discretization, both PQ and RPQ demonstrate significant performance improvements across all test sets. On the four English test sets, PQ ($M=16$) and RPQ ($M=32$) achieve average relative WER reductions of 23.9\% and 26.1\%, respectively, compared to the K-means baseline.
Similary, on the multilingual test set test-1h, PQ and RPQ achieve 20.0\% and 19.6\% relative reductions on CER, respectively. These results indicate that PQ and RPQ-based discretization methods offer notable advantages over the widely used K-means approach, reinforcing the idea that PQ and RPQ can retain more meaningful semantic information during discretization.

\begin{table*}[t]
    \centering
    \footnotesize
    \tabcolsep=3pt
    \caption{LibriSpeech WER (\%) on \{{\it dev, test}\}-\{{\it clean, other}\} and ML-SUPERB CER (\%) with different discretization methods. $V$ and $M$ denote the number of residual layers in RVQ and the number of subspaces in (R)PQ, respectively. The bitrate is computed accordingly.}
    {\color{black}
    \begin{tabular}{lc|cccccc|c}
    \toprule
    \multirow{2}[3]{*}{\textbf{Method}} & \multirow{2}[3]{*}{\textbf{Param.}} 
    & \multicolumn{5}{c}{\textbf{LibriSpeech WER (\%)}} & \multirow{2}[3]{*}{\textbf{ML-SUPERB CER (\%)}} & \multirow{2}[3]{*}{\textbf{Bitrate}}\\
    \cmidrule(lr){3-7}
     & & dev-clean & dev-other & test-clean & test-other & avg. & & \\
    \midrule
    \multicolumn{9}{l}{\textit{\textbf{Acoustic Tokens}}} \\
    \midrule
    EnCodec \cite{defossez2022high} & 8-level & - & - & - & - & 15.9 & 35.9 & 6000.0\\
    \midrule
    \multicolumn{9}{l}{\textit{\textbf{Semantic Tokens}}} \\
    \midrule
    K-means & - & 4.7 & 8.2 & 4.6 & 8.6 & 6.53 & 24.0 & 548.3\\
    ICA + K-means \cite{nakamura2025discrete} & - & - & - & - & - & - & 24.8 & 548.3\\
    Differentiable K-means \cite{onda2025differentiable} & - & - & - & 4.4 & 7.2 & 5.80 & - & 548.3\\
    \midrule
    \multirow{3}[2]{*}{RVQ \cite{shi2024mmm}} 
     & $V$=2 & - & - & - & - & 5.90 & 21.4 & 1096.6\\
     & $V$=4 & - & - & - & - & 6.10 & 21.5 & 2193.2\\
     & $V$=8 & - & - & - & - & 6.40 & 21.7 & 4386.3\\
    \midrule
    \multirow{5}[2]{*}{\textbf{PQ}} 
     & $M$=2 & 3.9 & 7.2 & 3.9 & 7.5 & 5.63 & 21.8 & 1096.6\\
     & $M$=4 & 3.7 & 6.9 & 3.7 & 7.2 & 5.38 & 20.0 & 2193.2\\
     & $M$=8 & {\bf 3.4} & 6.6 & 3.5 & 6.8 & 5.08 & 19.5 & 4386.3\\
     & $M$=16 & 3.5 & 6.5 & 3.5 & 6.9 & 5.10 & \textbf{19.2} & 8772.6\\
     & $M$=32 & 3.7 & 6.8 & 3.7 & 7.0 & 5.30 & 19.6 & 17545.3\\
    \midrule
    \multirow{5}[2]{*}{\textbf{RPQ}} 
     & $M$=2 & 3.9 & 7.5 & 4.0 & 7.9 & 5.83 & 21.4 & 1096.6\\
     & $M$=4 & 3.7 & 7.1 & 3.7 & 7.3 & 5.45 & 20.4 & 2193.2\\
     & $M$=8 & 3.6 & 6.7 & 3.6 & 7.0 & 5.23 & 19.9 & 4386.3\\
     & $M$=16 & 3.5 & 6.7 & \textbf{3.4} & 6.8 & 5.10 & 19.5 & 8772.6\\
     & $M$=32 & \textbf{3.4} & \textbf{6.4} & \textbf{3.4} & \textbf{6.6} & \textbf{4.95} & 19.3 & 17545.3\\
    \bottomrule
    \end{tabular}
    }
    \label{tab:RVQ}
\end{table*}

\begin{figure}
    \centering
    \includegraphics[width=0.7\linewidth]{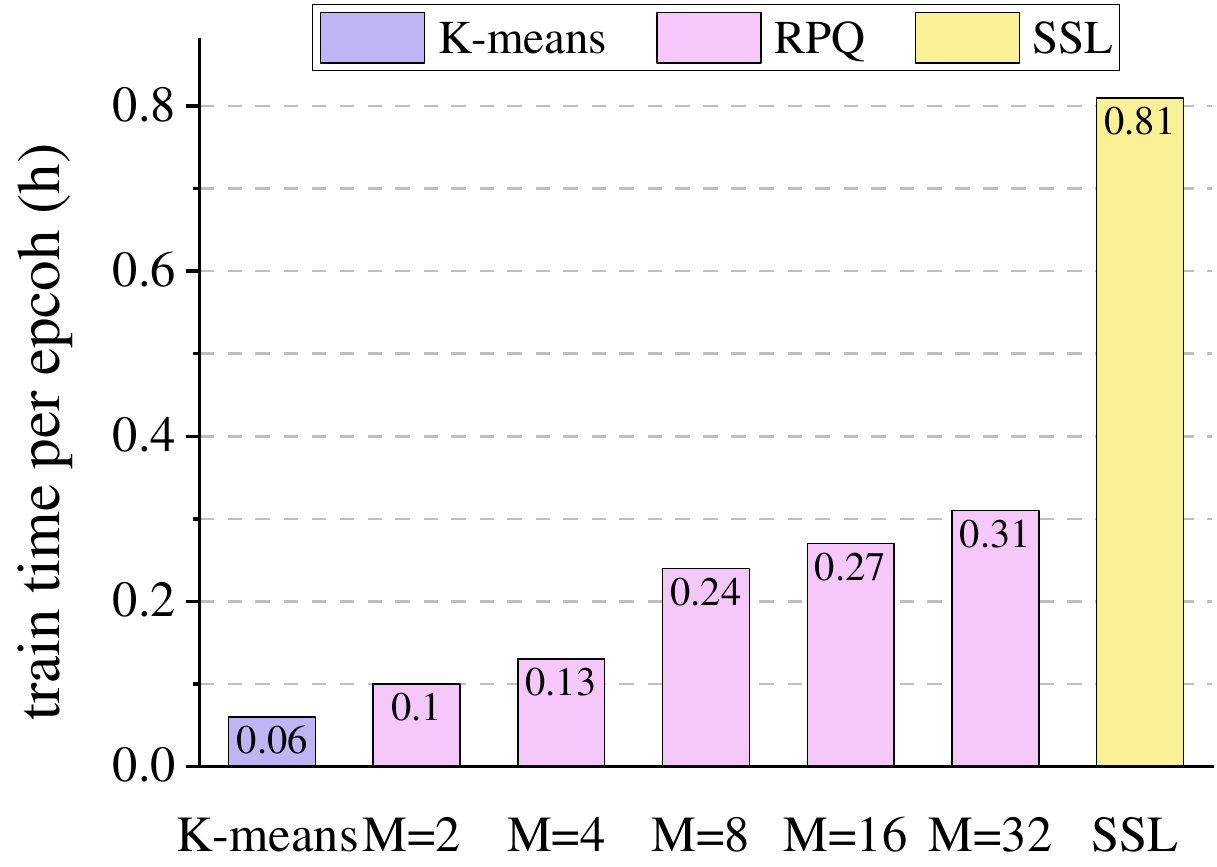}
    \caption{Comparison of training time per epoch when using K-means discrete unit / RPQ discrete unit / continuous SSL representation.}
    \label{fig:time}
\end{figure}

We further analyze the impact of the parameter $M$. As $M$ increases, both PQ and RPQ exhibit a steady performance improvement. For example, on the test-clean test set, when $M$ increases from 2 to 16, PQ’s WER decreases from 3.9\% to 3.5\%, achieving a relative improvement of 10.3\%. RPQ reaches its best performance at $M=32$, with a WER of 3.4\%, reflecting a 15.0\% improvement compared to 4.0\% at $M=2$. Comparing RPQ and PQ on English test sets, PQ outperforms RPQ when $M \leq 8$, while RPQ surpasses PQ at $M=16$ and $M=32$. This suggests that PQ achieves better quantization when the number of subspaces is small, whereas RPQ's random partitioning becomes advantageous when more subspaces are utilized. However, on the multilingual test set test-1h, PQ consistently yields better results.

{
\color{black}
For our proposed RPQ method, due to the stochastic nature of subspace partitioning, we additionally assess its stability by averaging results over multiple random seeds. 
All corresponding results are reported in the last row of \autoref{tab:main}, denoted as $mean \pm std$, where both statistics are computed across four runs with different random seeds.
The observed variance is consistently small, demonstrating that RPQ is robust to random initialization and partitioning.
}

\textcolor{black}{Compared to continuous representations, RPQ ($M$=32) achieves an average relative reduction of 3.9\% in WER on LibriSpeech and 11.1\% in CER on ML-SUPERB. These results suggest that RPQ substantially enhances the performance of discrete representations. Although discretization may introduce certain information loss, it can also serve as a form of ``information filtering,'' reducing irrelevant variability in the original continuous representations, such as para-linguistic factors (e.g., speaker characteristics and emotion) and non-linguistic factors (e.g., background noise). In particular, discretization-based speech modeling methods such as the proposed RPQ encourage the extraction of information that is most relevant to linguistic structure, leading to a representation space that is more focused on language-related patterns for subsequent ASR model training. Similar observations have also been discussed in prior work on the role of discrete representations in spoken language modeling~\cite{9864610}. Furthermore, the results indicate that the performance improvement of PQ and RPQ is more pronounced on ML-SUPERB.}
This improvement arises because discretization inherently compresses the information in continuous SSL representations. When the pretrained SSL model used for feature extraction is not well aligned with the target downstream task (i.e., WavLM Large was pretrained exclusively on English data and lacks exposure to multilingual knowledge), its learned representations tend to be biased toward English speech. Consequently, features extracted from multilingual speech exhibit a more dispersed distribution, necessitating the retention of additional information. In this context, K-means retains limited useful information when extracting discrete tokens, whereas PQ and RPQ, leveraging multiple codebooks, preserves more information relevant to the downstream task.

Fig. \ref{fig:time} demonstrates that both the baseline K-means and the proposed RPQ method achieve significantly higher training efficiency compared to the continuous SSL representation. Among discrete methods, RPQ incurs a moderate increase in training time as $M$ grows but remains much more efficient than SSL, with training time at $M = 32$ reaching only 38\% of that required by the continuous representation. These results highlight RPQ as a competitive alternative to K-means, balancing computational efficiency and discrete representation quality.

\textcolor{black}{\subsection{Comparison with State-of-the-Art Methods}}

\textcolor{black}{To further validate the effectiveness of our proposed method, we present a comparison with different state-of-the-art discretization methods evaluated on LibriSpeech (WER) and ML-SUPERB (CER) in \autoref{tab:RVQ}. Specifically, ICA + K-means \cite{nakamura2025discrete} applies independent component analysis preprocessing before clustering. However, it fails to show a performance advantage over the baseline k-means method. Differentiable K-means \cite{onda2025differentiable} enables the joint optimization of tokenization and downstream tasks, thus showing a significant performance advantage over the baseline on the LibriSpeech dataset. Additionally, we also report results using 8-level acoustic tokens from EnCodec \cite{defossez2022high} and a hierarchical quantization approach that recursively quantizes residuals, which is called RVQ \cite{shi2024mmm}. Notably, EnCodec performs substantially
worse than the K-means baseline, suggesting that acoustic
tokens are less suited for discrete speech recognition tasks. In contrast, both PQ and RPQ consistently outperform the baselines across a range of configurations, with even their weakest settings exceeding most of the compared baselines. This highlights the effectiveness of PQ and RPQ in discretizing SSL representations without requiring techniques like feature pre-processing or task-specific online optimization, making them more applicable to diverse scenarios.}

\textcolor{black}{Besides, we also report the bitrate of each discretization methods in \autoref{tab:RVQ}. For a fair comparison, all reported bitrates are raw values obtained without de-duplication and subword modeling. The proposed method employs a multi-codebook encoding strategy, which leads to a higher bitrate compared with the single-codebook k-means baselines. Nevertheless, its bitrate remains substantially lower than that of the original waveform (256 kbps) and the continuous SSL features (819.2 kbps), which are both calculated assuming 16-bit representations. This shows that our quantization approach achieves a favorable balance between representational capability and storage or transmission efficiency.}

\begin{table*}[htbp]
    \centering
    \caption{Performance under different ratios of $\alpha$.}
      \begin{tabular}{cccc|cccccccccc}
      \toprule
      \multirow{2}[4]{*}{\textbf{Method}} & \multirow{2}[4]{*}{$\boldsymbol{\alpha}$(\%)}& \multirow{2}[4]{*}{$\boldsymbol{\hat{\rho}}$(\%)}& \multirow{2}[4]{*}{$\boldsymbol{\rho}$(\%)} & \multicolumn{2}{c}{\textbf{test-clean}} & \multicolumn{2}{c}{\textbf{test-other}} & \multicolumn{2}{c}{\textbf{dev-clean}} & \multicolumn{2}{c}{\textbf{dev-other}} & \multicolumn{2}{c}{\textbf{test-1h}} \\
      \cmidrule(lr){5-6} \cmidrule(lr){7-8} \cmidrule(lr){9-10} \cmidrule(lr){11-12}\cmidrule(lr){13-14}     &&&      & CER   & WER   & CER   & WER   & CER   & WER   & CER   & WER   & CER   & WER \\
      \midrule
          & 6.25&\textcolor{black}{3.23}&\textcolor{black}{5.40}& 1.1   & 3.5   & 2.4   & 6.8   & 1.1   & 3.6   & \textbf{2.4} & 6.5   & 19.5  & 59.5 \\
          & 12.5&\textcolor{black}{6.67}&\textcolor{black}{13.03}& \textbf{1.0} & \textbf{3.4} & \textbf{2.3} & \textbf{6.6} & \textbf{1.0} & \textbf{3.4} & \textbf{2.4} & \textbf{6.4} & \textbf{19.3} & \textbf{59.3} \\
          & 25&\textcolor{black}{14.29}&\textcolor{black}{25.07}& \textbf{1.0} & 3.5   & 2.4   & 6.8   & 1.1   & \textbf{3.4} & \textbf{2.4} & 6.5   & 19.4  & \textbf{59.3} \\
        \textbf{RPQ} & 37.5&\textcolor{black}{23.08}&\textcolor{black}{29.79}& \textbf{1.0} & 3.5   & 2.4   & 6.9   & 1.1   & \textbf{3.4} & \textbf{2.4} & 6.5   & 19.5  & 59.6 \\
        ($M$=32, & 50&\textcolor{black}{33.33}&\textcolor{black}{32.56}& \textbf{1.0} & \textbf{3.4} & 2.4   & 6.8   & 1.1   & 3.5   & \textbf{2.4} & 6.5   & 19.7  & 59.7 \\
        $k$=2000) & 62.5&\textcolor{black}{45.45}&\textcolor{black}{40.39}& 1.1   & 3.5   & 2.4   & 6.9   & 1.1   & 3.5   & 2.5   & 6.6   & 19.6  & 59.9 \\
            & 75&\textcolor{black}{60}&\textcolor{black}{41.34}& 1.1   & 3.5   & 2.5   & 6.9   & 1.1   & 3.5   & 2.5   & 6.6   & 19.8  & 60.0  \\
            & 87.5&\textcolor{black}{77.78}&\textcolor{black}{44.34}& 1.2   & 3.7   & 2.5   & 7.2   & 1.2   & 3.7   & 2.6   & 6.8   & 19.8  & 60.2 \\
            & 100&\textcolor{black}{100.00}&\textcolor{black}{100.00}& 1.3   & 4.2   & 3.0   & 7.9   & 1.4   & 4.2   & 3.0   & 7.8   & 22.0  & 64.1 \\
    \bottomrule
      \end{tabular}
    \label{tab:alpha}
\end{table*}


\subsection{Analysis of Parameter \texorpdfstring{$\alpha$}{alpha}}

\begin{table*}[htbp]
    \centering
    \caption{Comparative experimental results of the datasets LibriSpeech, ML-SUPERB, and WenetSpeech using K-means discretization, PRQ discretization, and continuous SSL representation.}
      \begin{tabular}{l|cccc|cc||c}
      \toprule
      \multirow{2}[2]{*}{\textbf{Dataset}} & \multirow{2}[2]{*}{\textbf{SSL Model}} & \multirow{2}[2]{*}{\textbf{Language}} & \multirow{2}[2]{*}{\textbf{Metric}} & \multirow{2}[2]{*}{\textbf{Evaluation Sets}} & \multicolumn{3}{c}{\textbf{Results}} \\ &  &   &   &   & K-means & RPQ & Continuous SSL \\
      \midrule
      LibriSpeech & WavLM & EN & WER  & \{dev,test\}-\{clean,other\} & 3.8 / 6.7 / 3.9 / 7.0 & 3.7 / 6.2 / 3.6 / 6.4 & \textbf{3.2 / 5.3 / 3.1 / 5.5} \\
      ML-SUPERB & WavLM & 143  & CER  & test-1h & 25.2 & 19.1 & \textbf{18.9} \\
      ML-SUPERB & Data2vec2 & 143 & CER & test-1h & 35.9 & \textbf{22.4} & 24.6 \\
      WenetSpeech & Data2vec2 & CH & CER & test-net & 16.0 & 14.7 & \textbf{13.6} \\
      \bottomrule
      \end{tabular}
    \label{tab:other}
  \end{table*}

From the relationship between $\varepsilon _{\text{RPQ}}$ and $\varepsilon _{\text{kms}}$ in Eq. \eqref{eq.RPQ_kms}, it is evident that the correlation coefficient $\rho$ is closely related to the performance of the RPQ discretization method. \textcolor{black}{In our analysis, the theoretical correlation is given by $\hat{\rho} =\frac{\alpha}{2-\alpha}$. To further validate this, we supplement the measured values of $\rho$ under different $\alpha$ by calculating the representation similarity between sub-quantizer outputs. Specifically, we calculate the centered kernel alignment (CKA) \cite{kornblith2019similarity} score between the k-means centroid matrices of all sub-quantizers. The CKA quantitatively captures the degree of alignment between the learned subspaces, providing an empirical estimate of $\rho$. As illustrated in \autoref{tab:alpha}, both the theoretical $\hat{\rho}$ and measured values $\rho$ are positively correlated with the dimension selection ratio $\alpha$, highlighting the importance of choosing an appropriate $\alpha$ value.}
To further analyze the impact of $\alpha$ on performance, we conduct experiments using RPQ discretization ($M=32$, $k=2000$) while keeping $\alpha$ as the sole variable. The experimental results are presented in \autoref{tab:alpha}. As shown in the table, RPQ achieves the best performance across all four LibriSpeech test sets and the test-1h set when $\alpha$ is set to 12.5\%. However, as $\alpha$ continues to increase, performance gradually deteriorates. Notably, when $\alpha$ reaches 100\%, the WER on test-other and the CER on test-1h significantly rise to 7.9\% and 22.0\%, respectively.This suggests that an excessively high $\alpha$ results in an overly large $\rho$. According to the inference from Eq. \eqref{eq.RPQ_kms}, an excessively high $\rho$ diminishes the randomness in RPQ, causing its estimation error to approach the upper bound $\varepsilon_{\text{kms}}$. From a model learning perspective, excessive information redundancy in this case leads to overfitting. Conversely, if $\rho$ is too small, the performance of individual K-means models degrades, leading to a higher $\varepsilon_{\text{kms}}$. Overall, a moderate $\alpha$ effectively enhances recognition performance. For example, when $M=32$ and $k=2000$, the optimal $\alpha$ is around 12.5\%, while excessively high or low values tend to degrade performance.


\begin{figure}[t]
	\centering
	\subfigure[WER of test-other\label{fig:test-other}]{
		\includegraphics[width=0.45\linewidth]{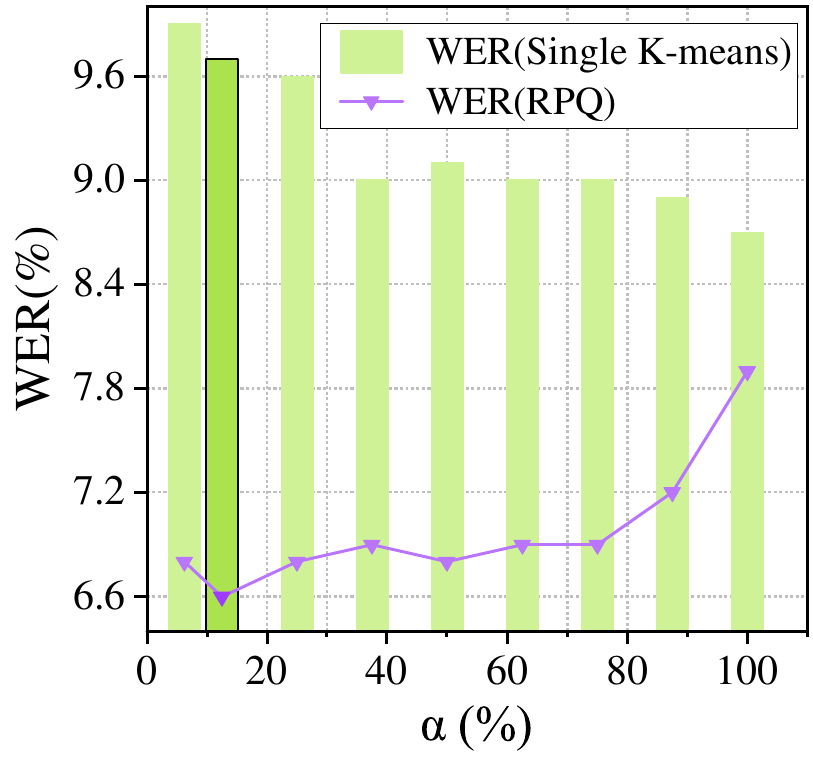}
	}
	\hspace{-10pt}
	\subfigure[CER of test-1h\label{fig:test-1h}]{
		\includegraphics[width=0.45\linewidth]{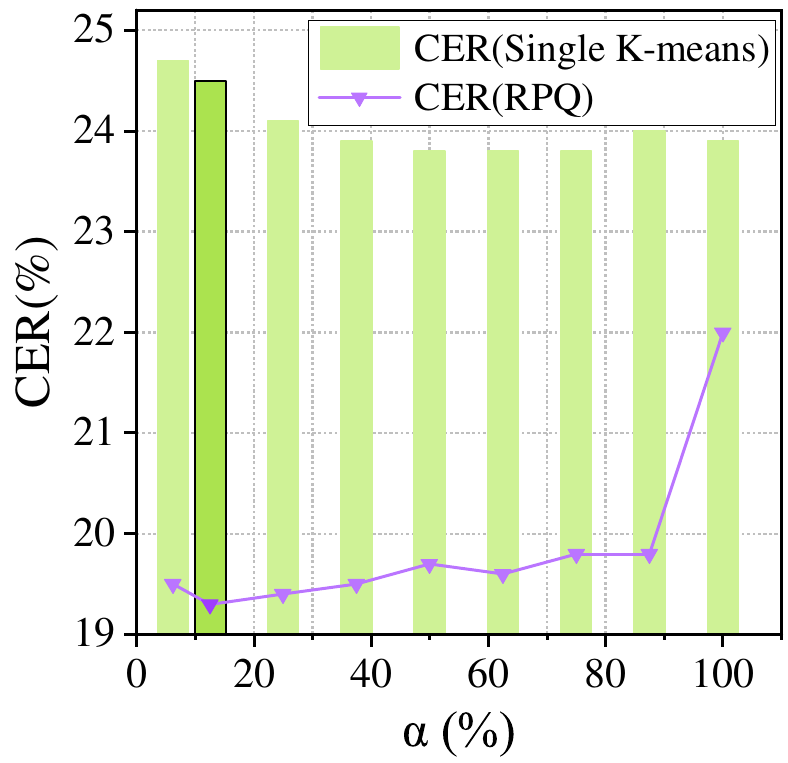}
	}
    \vspace{-5pt}
	\caption{Performance of the test sets test-other and test-1h when using RPQ and Single K-means discretization at different ratios $\alpha$. The bar graph is the result of Single K-means, and the line graph is the result of RPQ. The black border on the bar graph indicates that the experimental result of RPQ is optimal at this time.\label{fig:alpha}}
	\label{fig4}
\end{figure}

To better illustrate the impact of different $\alpha$ values on $\varepsilon_{\text{kms}}$ and $\varepsilon_{\text{RPQ}}$, we conduct a set of experiments using a single K-means model for discretization under varying $\alpha$ values. For clarity, we refer to this experiment as \textit{Single K-means}. Since directly computing the quantization error $\varepsilon_{\text{kms}}$ is challenging, we use WER and CER from the \textit{Single K-means} experiment as a proxy. The lower WER or CER indicates smaller $\varepsilon_{\text{kms}}$ and better recognition performance. The implementation of the Single K-means experiment is as follows: taking $\alpha = 25\%$ as an example, given a continuous speech representation of dimensionality $D = 1024$, a subset of $\alpha \times D = 256$ dimensions is randomly selected to form a subvector. This subvector is then quantized using a single K-means model to obtain discrete speech representations. It is important to note that the discrete token sequences obtained from the \textit{Single K-means} experiment differ from those in the baseline model. In the baseline model, quantization is applied to the full $D$-dimensional continuous representation, whereas in \textit{Single K-means}, the discrete tokens are derived from a subvector of $\alpha \times D$ dimensions. 

\begin{table*}[htbp]
    \centering
    \caption{Experiment on the impact of deduplication and BPE operations on experimental results, where "De-dup" and "Src\_bpe" indicate the deduplication step and BPE processing of the input discrete token sequences, respectively.}
      \begin{tabular}{ccc|cccccc}
      \toprule
      \multirow{2}[4]{*}{\textbf{Method}} & \multirow{2}[4]{*}{\textbf{De-dup}} & \multirow{2}[4]{*}{\textbf{Src\_bpe}} & \multicolumn{2}{c}{\textbf{test-clean}} & \multicolumn{2}{c}{\textbf{test-other}} & \multicolumn{2}{c}{\textbf{test-1h}} \\
      \cmidrule(lr){4-5} \cmidrule(lr){6-7} \cmidrule(lr){8-9}          &       &       & CER   & WER   & CER   & WER   & CER   & WER \\
      \midrule
      K-means & $\checkmark$  & 3000  & 1.5   & 4.6   & 3.3   & 8.6   & 24.0  & 68.3 \\
      K-means & $\times$  & $\times$  & 1.5   & 4.6   & 3.3   & 8.7   & 23.9  & 67.9  \\
      \bottomrule
      \end{tabular}%
    \label{tab:bpe}%
\end{table*}%

Fig. \ref{fig:alpha} illustrates the variation of WER on the test-other set and CER on the test-1h set (bar chart) as $\alpha$ increases. It can be observed that as $\alpha$ grows from 6.25\% to 100\%, the WER and CER of Single K-means exhibit a gradual decline. This trend indicates that with a larger $\alpha$, Single K-means can utilize a more complete representation, leading to a reduction in quantization error and an improvement in recognition performance. Additionally, Fig. \ref{fig:alpha} also presents the experimental results of RPQ (line chart). Taking the test-other results in the left plot as an example, it is more apparent that as $\alpha$ increases, the WER of RPQ first decreases and then increases, reaching its optimal performance at $\alpha = 12.5\%$.

\subsection{Experimental Comparison Results on Other Datasets}

To evaluate RPQ on different datasets, we conducted experiments on LibriSpeech, ML-SUPERB, and the Chinese dataset WenetSpeech, comparing K-means discretization, RPQ discretization, and continuous representations. As shown in \autoref{tab:other}, RPQ outperformed K-means across all datasets. On LibriSpeech, RPQ reduced WER by 2.6\% and 7.5\% relative to K-means on dev-clean and dev-other, respectively. On ML-SUPERB, RPQ achieved relative CER reductions of 24.2\% and 37.6\% with WavLM and Data2vec representations, nearly matching continuous representations and even surpassing them by 8.9\% with Data2vec. On WenetSpeech, RPQ improved CER by 8.1\% over K-means and closely approached continuous representations.  Overall, RPQ consistently outperformed K-means across datasets and models, often rivaling or exceeding continuous representations, demonstrating its effectiveness in ASR.

\subsection{Verification Experiments}

Since PQ and RPQ discretization partition speech representations into multiple subspaces, ensuring strict alignment among multiple discrete representations during subsequent fusion is crucial. Therefore, no deduplication or BPE subword modeling was applied to the discrete token sequences. To verify that the performance improvement of PQ and RPQ over the baseline is not merely due to skipping the deduplication and BPE steps, and to investigate the impact of these steps on experimental results, this section presents a comparative study on the mixed dataset. Specifically, we examine whether K-means discretization in the baseline approach applies deduplication and BPE to the input discrete token sequences. As shown in \autoref{tab:bpe}, the CER and WER results on the test-clean, test-other, and test-1h sets are highly similar, indicating that deduplication and BPE have minimal impact on the performance of ASR.

\begin{table}[t]
    \centering
    \caption{PQ experiment under different warmup\_steps, where \textbf{peak} refers to the epoch at which the learning rate curve reaches its maximum during training.
    }
      \begin{tabular}{ccc|cccc}
      \toprule
      \multirow{2}[4]{*}{\textbf{Method}} & \multirow{2}[4]{*}{\textbf{Warmup\_steps}} & \multirow{2}[4]{*}{\textbf{Peak}} & \multicolumn{2}{c}{\textbf{dev-clean}} & \multicolumn{2}{c}{\textbf{dev-other}} \\
      \cmidrule(lr){4-5} \cmidrule(lr){6-7}  &  &  & CER  & WER   & CER   & WER \\
      \midrule
      K-means & 5k    & \textbf{16} & 1.6   & 4.7   & 3.3   & 8.2 \\
      \midrule
      \textbf{PQ} & 5k    & 3     & 1.9   & 5.1   & 3.4   & 8.5 \\
      ($M$=32, & 10k   & 6     & 1.2   & 3.8   & 2.6   & \textbf{6.8} \\
      $k$=2000) & 30k   & \textbf{16} & \textbf{1.1} & \textbf{3.7} & \textbf{2.5} & \textbf{6.8} \\
      \bottomrule
      \end{tabular}%
    \label{tab:warmup}%
\end{table}%

\begin{table}[t]
	\caption{\color{black} Experiment on speech reconstruction capability of different discretization methods. 
    } 
	\centering
	\label{tab:sr}
    \setlength{\tabcolsep}{0.5mm}{\color{black}
	\begin{tabular}{c|ccccc}
		\toprule
		{\bf Method}
            &{\bf UTMOS}&{\bf PESQ-WB}&{\bf STOI}&{\bf SIM (\%)}&{\bf WER (\%)}\\
          \midrule
		
		Ground Truth&4.09&-&-&-&3.2  \\
        \midrule
        \midrule
        {Continuous SSL}  &3.47&1.52&0.87&81.1&3.6 \\
        \midrule
		{K-means}&3.10&1.08&0.64&59.0&4.7  \\
            {\bf PQ} ($M$=32, $k$=2000)  &{\bf 3.37}&{\bf 1.14}&{\bf 0.74}&{\bf 65.8}&4.4 \\
        {\bf RPQ} ($M$=32, $k$=2000)&3.29&1.12&0.72&63.7&{\bf 4.2}  \\
		\bottomrule
	\end{tabular}}
\end{table}

In PQ and RPQ discretization experiments, handling multiple discrete unit sequences simultaneously increases the number of model updates per batch. This causes the learning rate curve to peak earlier, leading to worse performance. To align the peak position of the learning rate curve more closely, we adjust the warmup\_steps parameter in the PQ experiment. As shown in \autoref{tab:warmup}, the best performance is achieved with warmup\_steps=30k, where the learning rate peak occurs around epoch 16, similar to the baseline K-means experiment. In contrast, with warmup\_steps=5k or 10k, the peak appears too early, causing the learning rate to rise rapidly before the model has sufficiently converged, leading to unstable training.
{\color{black}
\subsection{Exploration Beyond the ASR Task}

To further evaluate the proposed quantization methods in terms of semantic preservation and information fidelity, we conduct a speech reconstruction experiment. Specifically, for each type of feature, we train a Diffusion Transformer \cite{dit} with a flow matching \cite{fm} loss for speech reconstruction, and assess the results using UTMOS \cite{utmos} and PESQ-WB \cite{PESQ} for perceptual quality, STOI \cite{STOI} for intelligibility, SIM calculated with {\it wespaker} toolkit \cite{wespeaker} for speaker similarity, and WER computed using {\it Whisper-small.en} \cite{whisper} to measure the preservation of linguistic information.

The results, summarized in \autoref{tab:sr}, show that both PQ and RPQ substantially outperform the classical K-means baseline across all perceptual and intelligibility metrics. In particular, PQ achieves the highest overall reconstruction quality, while RPQ attains the lowest WER, demonstrating its superior capability in preserving linguistic content through diversified subspace sampling. These findings confirm that the proposed PQ and RPQ methods retain richer semantic information in discrete speech representations, effectively narrowing the gap between continuous and discrete modeling. We further show some reconstructed speech samples on our supplementary page\footnote{\href{https://aisaka0v0.github.io/rpq_supplementary_page/}{\color{black} https://aisaka0v0.github.io/rpq\_supplementary\_page/}}.
}

\section{Conclusion} \label{conclusion}

This paper proposes two methods for discretizing self-supervised speech representations: Product Quantization (PQ) and Random Product Quantization (RPQ). Both approaches are motivated by the goal of preserving more meaningful semantic information during the discretization process, thereby narrowing the performance gap between discrete and continuous speech recognition systems. PQ decomposes continuous representations into multiple low-dimensional sub-vectors, each of which is quantized independently. The resulting discrete codes are then combined and used as input to downstream models. RPQ follows a similar procedure but introduces randomness by repeatedly selecting subsets of dimensions at a fixed ratio from the feature space to construct low-dimensional sub-vectors, which are then independently quantized. While PQ enables discrete representations to retain information from multiple vector subspaces, RPQ increases the diversity among these subspaces, allowing the discrete representations to capture a richer range of speech characteristics. In addition, we present a rigorous theoretical analysis of the quantization error associated with RPQ, providing formal guarantees and insights into its performance. Experimental results show that both PQ and RPQ significantly outperform classical K-means and other existing discretization techniques. As a future direction, we plan to explore integrating PQ and RPQ with large language models, aiming to further enhance the capability of discrete speech representations in broader multimodal tasks.

\small
\bibliographystyle{IEEEtran}
\bibliography{Reference}

\end{document}